\documentclass[final,5p,times,twocolumn]{elsarticle}
\usepackage[T1]{fontenc}
\usepackage{bm}
\usepackage{amsfonts}
\usepackage{graphicx}
\usepackage{slashed}
\usepackage{color}
\usepackage{cancel}
\usepackage{hyperref}
\usepackage{amsthm}
\usepackage{amssymb}
\usepackage{amsmath}
\usepackage{xparse}
\usepackage{xspace}
\usepackage{mathtools}
\usepackage{etextools}
\usepackage{ifthen}
\usepackage{empheq}
\usepackage[most]{tcolorbox}
\usepackage{flushend}

\newcommand{\COM}[2]{\ensuremath{\left[{#1},{#2}\right]}\xspace}
\newcommand{\Order}[1]{\mathcal{O}\!\left(#1\right)}
\DeclarePairedDelimiter{\MbBraces}{\{\!\{} {\}\!\}}
\newcommand{\PB}[2]{\left\{#1,#2\right\}}
\newcommand{\MB}[2]{\MbBraces{#1,#2}}
\newcommand{\ud}{\mathrm{d}}
\newcommand{\uW}{\mathbf{W}}
\newcommand{\uF}{\mathbf{F}}
\newcommand{\uA}{\mathbf{A}}
\newcommand{\uB}{\mathbf{B}}

\newcommand{\uE}{\mathbf{E}}
\newcommand{\uG}{\mathbf{G}}
\newcommand{\uH}{\mathbf{H}}
\newcommand{\uK}{\mathbf{K}}
\newcommand{\uX}{\mathbf{X}}

\newcommand{\ui}{\mathrm{i}}
\newcommand{\Id}{\mathbf{1}_4}
\newcommand{\IdN}{\mathbf{1}_N}
\newcommand{\Zd}{\mathbf{0}_4}

\NewDocumentCommand{\PD}{o m m}{%
  \IfNoValueTF{#1}
    {\ensuremath{\frac{\partial #2}{\partial #3}}}%
    {\ensuremath{\frac{\partial^{#1} #2}{\partial #3^{#1}}}}%
  \xspace
}

\usepackage{amsmath,amssymb,bm}
\usepackage[normalem]{ulem}
\usepackage{needspace}

\hypersetup{hidelinks}
\journal{Physics Letters A}

\begin{document}

\begin{frontmatter}

\title{
Quantum Mechanics as a Relativistic Statistical Theory: \\
A Mass-Shell Factorisation Approach
}

\author[1]{Mark Everitt\corref{cor1}}
\cortext[cor1]{Corresponding author}
\ead{m.j.everitt@physics.org}
\address[1]{Quantalytics, Spinney Hill Drive, Loughborough, LE11 3LB, UK}

\begin{abstract}
Spinor structure is usually introduced as part of relativistic quantum theory. 
We show that its underlying algebra can instead emerge without quantisation from a covariant statistical description of a massive relativistic particle subject to an explicit finite-dimensional linear factorisation of the complete massive mass-shell constraint motivated by requiring the theory retain both sheets of the massive mass shell.
A finite-dimensional factorisation of the complete mass-shell constraint, linear in all four momentum components, forces a Clifford algebra. 
In (3+1) dimensions its minimal complex representation is four-dimensional, while each on-shell factor selects a rank-two sector; in this representation statistical completeness requires an arbitrary $2\times2$ matrix distribution within that sector before quantisation. 
Projecting the Weyl-ordered matrix Liouvillian gives relativistic transport while preserving all internal populations and coherences. 
For universal deformations on flat canonical phase space satisfying the stated covariance, matrix-neutrality, homogeneity and associativity assumptions, the resulting product is the matrix Weyl--Moyal product. 
Its parameter has dimensions of action and, when identified with $\hbar$, supplies the scale of physical spin-$\tfrac12$. 
Requiring all-order constraint preservation gives a vanishing star commutator, while preservation of the selected star-spectral sector yields the two-sided Dirac--Wigner equations. 
Our central conclusion is therefore that relativistic mass-shell factorisation supplies spinor structure, while completion of the constraint-preservation hierarchy within the stated deformation class leads to the Weyl--Moyal phase-space formulation of quantum mechanics.
\end{abstract}

\begin{keyword}
	Relativistic statistical mechanics \sep 
	covariant classical phase space    \sep 
	classical spinor structure 		   \sep 
	Clifford algebra                   \sep 
	mass-shell factorisation           \sep 
	deformation quantisation           \sep 
	Dirac--Wigner equations
\end{keyword}

\end{frontmatter}
\setcounter{secnumdepth}{0}
\tableofcontents
\newtheorem{ansatz}{Ansatz}
\section{Introduction}
Spin-$1/2$ is normally introduced into relativistic quantum mechanics through a spinor state space. 
We show the underlying spinor structure can instead follow from compatibility between relativistic kinematics and statistical phase-space dynamics. 
The resulting constraint-preservation hierarchy, together with the deformation assumptions specified in Part II, recovers the Weyl--Moyal phase-space equations.
This study was motivated by the view that any physical theory should allow statistical descriptions of ensembles without selecting a preferred observer.
This makes the covariant form of Liouville's equation a natural starting point, although its phase-space generator need not have the interpretation
of a coordinate-time energy~\cite{hakim2011introduction}. 
Canonical interacting particle models are also restricted by the Currie--Jordan--Sudarshan no-interaction theorem, which shows that, for a finite number of classical point particles with canonical coordinates whose positions transform covariantly, there is no non-trivial Hamiltonian interaction compatible with the full Poincar\'e symmetry (the fundamental symmetries of special relativity)~\cite{RevModPhys.35.350}. 
Phase-space formulations hold the promise that quantum mechanics naturally emerges as a deformation quantisation of classical physics, replacing the Poisson bracket with the Moyal bracket and conventional distribution multiplication with the star product~\cite{GROENEWOLD1946405,Moyal_1949}. 
However, a persistent gap in this approach has been the difficulty of naturally introducing spin, since quantum mechanical spin-$1 \over 2$ has no direct analogue in classical mechanics. 
Previous work has introduced spin phase-space representations through $\mathrm{SU}(N)$ Wigner functions formulated in terms of a displaced-parity operator~\cite{PhysRevA.99.012115}.
More recently, Ji and Piasecki mapped the relativistic Wigner operator, through its Clifford decomposition, to a Koopman phase-space spinor with a classical Liouville limit~\cite{sym18061018}.

This paper is organised in two parts. In the first part we consider the free particle and present a construction the  question of how the Clifford algebra and its rank-two spinor structure follow directly from first-order mass-shell factorisation, before quantisation.
The argument developed below revisits Dirac's original derivation, but reformulates it in a classical phase-space language: 
the relativistic Liouville equation replaces Schr\"odinger's equation as the dynamical starting point, and spin degrees of freedom are encoded via a Clifford algebra-valued distribution. 
For a relativistic extension of Schr\"odinger's equation, Dirac argued that 
(i) it should make use of the mass shell, 
(ii) the dynamical equation, like the Schr\"odinger equation, should be first-order in time but also, inspired by issues with the Klein-Gordon equation, that the Hamiltonian should be first-order in space, and 
(iii) the four-momentum components should be replaced with their operator counterparts~\cite{Dirac28}. 
Unlike Dirac's argument, we will use the Liouville equation as the basis for the discussion, but apart from this, and the fact that no operator substitution is required, we impose the following three requirements: 
(a) the dynamics should be first order in spacetime and momentum derivatives 
(b) both sheets of the relativistic mass shell should be retained, and 
(c) the relativistic mass-shell relation should admit a factorisation that treats space and time on the same footing by being linear in both energy and momentum. 
These requirements lead naturally to a Clifford factorisation. 
In $3+1$ dimensions its minimal complex representation is four-dimensional, while its on-shell eigenspaces have rank two.
In this sense, the algebra of spin arises kinematically from the compatibility of relativistic mass-shell factorisation with a statistical phase-space description.
At each on-shell momentum the factor selects a two-dimensional subspace. 
A statistical ensemble that retains all information within this subspace is therefore naturally matrix valued before any attempt at quantisation.

In the second part we show that the phase-space formulation of relativistic quantum mechanics arises from completing the constraint-preservation hierarchy identified in the free-particle case. 
Rather than postulating a star product of known form, we begin with a general universal, local and unital bidifferential deformation of the matrix-valued phase-space algebra. 
Translation invariance and affine canonical covariance, together with constant-matrix neutrality, the absence of additional background structure and order-by-order homogeneity, restrict the deformation to a series in powers of the canonical Poisson operator. 
Associativity then determines the coefficients of this series, while the first-order preservation condition fixes its leading coefficient, yielding the matrix Weyl–Moyal product.
The first-order term also recovers the symmetrised matrix Liouvillian introduced in Part I with the benefit of hindsight. 
Dimensional consistency requires the deformation parameter to have units of action, and comparison with established quantum mechanics identifies it with $\hbar$. 
Requiring the constraint-preservation hierarchy to hold at every order produces the vanishing star-commutator equation, while preservation of the selected star-spectral sector gives the corresponding left and right stargenvalue equations. 
For the relativistic mass-shell factor, these reproduce the Dirac–Wigner equations. Within the stated assumptions, the product, transport equation and spectral constraints of quantum phase-space mechanics therefore emerge as successive stages of a single constraint-preservation argument.

Thus a classical relativistic mass-shell construction, supplemented by the stated factorisation, deformation and spectral assumptions, recovers the Weyl--Moyal and Dirac--Wigner phase-space equations.


\section{Part I -- Free particle argument for spinor structure}

We work on the extended phase space formed by the spacetime coordinates
$  
    x\,^\mu=(ct,\bm{x})
$ where Greek indices run from $0$ to $3$, repeated indices are summed, and we use the metric $\eta_{\mu\nu}=\operatorname{diag}(1,-1,-1,-1)$. 
Proper time $\tau$ is defined along any given worldline by $\ud s = c\,\ud\tau $ where the line element $\ud s^2 = \eta_{\mu\nu} \, \ud x\,^\mu \ud x\,^\nu = c^2 \ud t^2 - \ud\Vec{x}\,^2
$. 
The three-velocity is $v^i \equiv \ud x\,^i/\ud t$, and $\Vec{v}$ denotes the corresponding three-vector. 
The physical three-momentum is $\Vec{p}=\gamma m\Vec{v}$, where
$\gamma=(1-\Vec{v}\,^2/c^2)^{-1/2}$, and the corresponding
future-directed four-momentum is $p\,^\mu_+=(E_p/c,\Vec{p})$, with
$E_p=\sqrt{c^2\Vec{p}\,^2+m^2c^4}$. To retain the complete mass shell,
we extend the phase-space description to
$p^\mu_\sigma=(\sigma E_p/c,\Vec{p})$, where the mass-shell label $\sigma=\pm1$. The independent $(x\,^\mu,p_\mu)$ coordinates form the eight-dimensional \emph{``ambient''} phase spacetime. The mass-shell constraint selects two seven-dimensional hypersurfaces within it.

\begin{ansatz}[Flat phase spacetime]
\label{ans:flat-phase-spacetime}
We work on $3+1$-dimensional Minkowski phase spacetime with
canonical coordinates $(x^\mu,p_\mu)$, metric $\eta_{\mu\nu}=\operatorname{diag}(1,-1,-1,-1)$, and commuting
classical phase-space variables.
\end{ansatz}
The relativistic Liouville equation is defined in terms of $H(x,p)$, a scalar covariant phase-space generator (rather than necessarily a coordinate-time energy) and $W(x,p)$, a scalar ensemble distribution.  Collisionless evolution is expressed through the vanishing extended Poisson
bracket~\cite{MARSDEN198629,hakim2011introduction}
\begin{equation}
  \mathcal{L}_H(W)
  \equiv
  \PB{H}{W}
  \equiv 
  \PD{H}{p_\nu}
  \PD{W}{x^\nu}
  -
  \PD{W}{p_\nu}
  \PD{H}{x^\nu}
  =0.
  \label{eq:relL}
\end{equation}
Because this equation is formulated covariantly on extended phase space, we do not identify $H$ with the coordinate-time energy
$\gamma mc^2$. Instead, we begin from the invariant mass-shell
constraint
\begin{equation}
  p\,^\mu p_\mu=m^2c^2,
  \label{eq:mass-shell}
\end{equation}
Solving the constraint only for $p^0=E_{\bm p}/c$ would select the future-directed sheet. We instead seek a single Lorentz-covariant factor that is linear in all four components of $p^\mu$. 
\begin{ansatz}[Complete first-order mass-shell factorisation]
\label{ans:complete-mass-shell-factorisation}
We consider a free particle with $m>0$ and retain both sheets of the massive mass shell,
$$
    p^\mu p_\mu=m^2c^2,
    \qquad \operatorname{sgn}(p^0)=\pm1.
$$
We require the complete constraint to admit a single
finite-dimensional Lorentz-covariant expression
$$
    K(p)=A^\mu p_\mu
$$
that is linear in all four components of the commuting classical four-momentum. 
Its square is required to reproduce the quadratic form for arbitrary $p$:
\begin{equation}
    K(p)^2
      =A^\mu A^\nu p_\mu p_\nu
      =\left(p^\mu p_\mu\right)\IdN
      \qquad \text{for all }p .
    \label{eq:linear-factor-square}
\end{equation}
\end{ansatz}
Because $p_\mu p_\nu$ is symmetric, only the anticommutator of the
coefficient matrices contributes:
\begin{equation}
    K(p)^2
    =
    \frac{1}{2}
    \COM{A^\mu}{A^\nu}_+ p_\mu p_\nu .
\end{equation}
Comparison of the coefficients of $p_\mu p_\nu$ therefore requires
\begin{equation}
    \COM{A^\mu}{A^\nu}_+
    =
    2\eta^{\mu\nu}\IdN.
\end{equation}
Thus the coefficient matrices necessarily generate the Clifford algebra. 
\begin{ansatz}[Minimal Clifford representation]
\label{ans:minimal-clifford-representation}
We take the coefficient matrices to act in an irreducible complex representation of the Clifford algebra of minimal dimension. 
In $3+1$ dimensions this fixes $N=4$. 
Larger reducible representations would introduce additional degeneracies not required by the mass-shell factorisation.
\end{ansatz}
Writing $A^\mu\equiv\gamma^\mu$, the massive constraint factorises as
\begin{equation}
    \left(\gamma^\mu p_\mu-mc \Id\right)
    \Big(\gamma^\nu p_\nu+mc \Id\Big)
    =
    \left(p^\mu p_\mu-m^2c^2\right)\Id.  \label{eq:factorisation}
\end{equation}
On the mass shell, the product in Eq.~(6) vanishes, and each linear factor has a nontrivial kernel. Importantly, the components of $p\,^\mu$ remain commuting classical variables where no operator substitution has been made. 
The construction mirrors Dirac's algebraic factorisation, but it is carried out entirely on relativistic phase space. 
We therefore introduce the matrix-valued covariant phase-space generator
\begin{equation}
  \uH_{\mathrm{free}}(p)
  =c(\gamma\,^\mu p_\mu-mc \Id).
  \label{eq:matrix-generator}
\end{equation}
The linear factor acts in the minimal four-dimensional representation of the Clifford algebra. 
To formulate a statistical theory on this representation, we introduce a $4\times4$ matrix-valued covariant phase-space function $\uW(x,p)$, which we call the matrix ensemble distribution. 
The need for a general matrix distribution does not follow from the representation alone, rather it follows below when we require the statistical description to retain all information within the subspace selected by the factor constraint.
One extension of Eq.~\eqref{eq:relL} to matrix form is
\begin{equation}
  \mathcal{L}_{\uH}(\uW)
  \equiv
  \PD{\uH}{p_\nu}
  \PD{\uW}{x^\nu}
  -
  \PD{\uW}{p_\nu}
  \PD{\uH}{x^\nu}
  \label{MLE}
\end{equation}
which we note retains the ordering of the usual, but not necessary, expression of the Poisson bracket.
Under a Lorentz transformation $\Lambda$, the distribution transforms
as
\begin{equation} \label{eq:similarity_law}
  \uW(x,p)\longmapsto
  S(\Lambda)
  \uW(\Lambda^{-1}x,\Lambda^{-1}p)
  S^{-1}(\Lambda).
\end{equation}
Because the factors are matrix valued, their order matters. The bracket in Eq.~\eqref{MLE} is bilinear and antisymmetric, but it does not in general satisfy the Jacobi identity or the Leibniz rule expected of a Lie bracket. 
It should therefore be regarded as one possible ordered extension of Eq.~\eqref{eq:relL} rather than as a Poisson bracket on a matrix algebra.

Because $\uH(x,p)$ and $\uW(x,p)$ no longer commute there are alternative brackets that would reproduce the normal Poisson bracket under the circumstances that $\uH$ and $\uW$ commute. 
For example consider the symmetric (Weyl-ordered) Liouvillian 
\begin{align}
    &
    \mathcal{L}^{\mathcal{W}}_\uH(\uW) \equiv
   \label{extendedbracket} 
   \\  & \ \ \ \ \ \ \ \ \ \ \ \ \ \ 
    \nonumber
    \frac{1}{2}
    \left[
    \left(
	\PD{\uH}{p_\nu}
	\PD{\uW}{x^\nu}
	+
	\PD{\uW}{x^\nu}
	\PD{\uH}{p_\nu}
	\right)
    -
	\left(
	\PD{\uH}{x^\nu}
	\PD{\uW}{p_\nu}
	+
	\PD{\uW}{p_\nu}
	\PD{\uH}{x^\nu}
	\right)
    \right]
\end{align}
This ordering is introduced here as at this stage a guess and the mass-shell factorisation by itself does not determine it. 
In Part II, we find that physically important Ansatz such as translation invariance and linear symplectic covariance, together with constant-matrix neutrality, the absence of additional phase-space tensors and first-order homogeneity together selects the symmeterised Liouvillian from alternatives. 
At this stage we note that this is neither a Lie bracket nor a derivation. We will later find that it is nevertheless singled out by deformation quantisation, where we will see that it appears as the local $\Order{\hbar}$ term of the matrix star commutator. 
We do not yet impose the vanishing of the full matrix expression as the admissible classical matrix space should first be identified from the factor constraint.

Following~\cite{thaller1992dirac,BjorkenDrell1964RQM} we introduce the covariant mass-shell projectors
\begin{equation}
    \Pi_s(p)
    =
    \frac{1}{2}
    \left(
        \Id+s\frac{\gamma^\mu p_\mu}{mc}
    \right),
\end{equation}
Here $s=\pm 1$ labels the eigenspaces of $\gamma\,^\mu p_\mu/mc$ and should not be confused with the mass-shell label $\sigma=\operatorname{sgn}(p^0)$.
Note that~\cite{BjorkenDrell1964RQM} used the notation $\Lambda_\pm(p)$ which we have changed to avoid confusion with Lorentz transformations.
For any on-shell momentum $p_\sigma$, both $\Pi_+(p_\sigma)$ and $\Pi_-(p_\sigma)$ have rank
two. 
Moreover,
\begin{equation}
  \uH_{\mathrm{free}}(p_\sigma)\Pi_s(p_\sigma)
  =mc^2(s-1)\Pi_s(p_\sigma).
  \label{eq:projector-kernel}
\end{equation}
The zero-eigenvalue constraint therefore selects $s=+1$ on either
mass-shell sheet, while $\sigma$ distinguishes the two energy
branches.
These mass-shell projectors obey 
    $\Pi_s^2         = \Pi_s$, 
    $\Pi_s\Pi_r      = \delta_{sr}\Pi_s$,   
    and $\Pi_+ + \Pi_-   = \Id$
when $p^\mu p_\mu=m^2c^2$.

At each on-shell momentum, the factor consequently selects the rank-two subspace
\begin{equation}
    \mathcal{E}_{p_\sigma}
    \equiv
    \ker \uH_{\mathrm{free}}(p_\sigma)
    =
    \operatorname{im}\Pi_+(p_\sigma).
    \label{eq:constraint-subspace}
\end{equation}
We now impose:
\begin{ansatz}[Statistical completeness within the selected sector]
\label{ans:statistical-completeness}
At each on-shell momentum $p_\sigma$, the ensemble description is required to retain arbitrary populations and coherences within the selected subspace
$$
    \mathcal E_{p_\sigma}
       =\ker H_{\mathrm{free}}(p_\sigma)
       =\operatorname{im}\Pi_+(p_\sigma).
$$
Accordingly,
$$
    \uW_0=\Pi_+\uW_0\Pi_+,
$$
and the restriction of $\uW_0$ to $\mathcal E_{p_\sigma}$ is an arbitrary matrix-valued distribution
$$
    F_\sigma(x,p)\in M_2(\mathbb C).
$$
\end{ansatz}
In the standard Dirac representation, the projectors at rest are
\begin{equation}
    \Pi_+(p_+)
    =
    \begin{pmatrix}
        1_2 & 0_2\\
        0_2 & 0_2
    \end{pmatrix},
    \qquad
    \Pi_+(p_-)
    =
    \begin{pmatrix}
        0_2 & 0_2\\
        0_2 & 1_2
    \end{pmatrix}.
\end{equation}
The corresponding distributions therefore have the forms
\begin{equation}
  \hspace{-4pt}  
  \uW_0(x,p_+)
    =
    \begin{pmatrix}
        \uF_+(x,\bm p) & 0_2\\
        0_2 & 0_2
    \end{pmatrix},
    \uW_0(x,p_-)
    =
    \begin{pmatrix}
        0_2 & 0_2\\
        0_2 & \uF_-(x,\bm p)
    \end{pmatrix}
    \label{eq:projected-classical-ansatz}
\end{equation}
where $\uF_\sigma$ is an arbitrary $2\times2$ matrix distribution acting within the selected subspace.
The position of the nonzero $2\times2$ block depends on the mass-shell sheet and the chosen basis (note that for general nonzero momentum, however, the boost mixes the upper and lower components, so neither simple block form holds in the fixed Dirac basis).
Using $\uW_0=\Pi_{+}\uW_0\Pi_{+}$ together with $\uH_{\mathrm{free}}\Pi_{+} =\Pi_{+}\uH_{\mathrm{free}}=\Zd$, we obtain
\begin{equation}
    \uH_{\mathrm{free}}\uW_0
    =\Zd
    =\uW_0\uH_{\mathrm{free}} \text{\ \  and \ \  }
    \COM{\uH_{\mathrm{free}}}{\uW_0}=\Zd.
    \label{eq:classical-commutator-constraint}
\end{equation}
This commutator is a classical compatibility condition between the distribution and the factor constraint and introduces no scale of action. If both eigenspaces of the factor are retained, the weaker condition is $\uW_0=\sum_s\Pi_s\uW_0\Pi_s$, which is equivalent on shell to $\COM{\uH_{\mathrm{free}}}{\uW_0}=\Zd$. The zero-eigenvalue constraint selects the $\Pi_+$ term, while the mass-shell label $\sigma$ continues to distinguish the two energy sheets.

In the positive-energy rest frame where $p^\mu=(mc,\Vec{0})$, in the standard Dirac representation, we see
$$
        \Pi_+(mc,\mathbf{0})
        =
        \begin{pmatrix}
        1_2 & 0\\
        0 & 0
        \end{pmatrix}
        \text{\ \  and \ \ }
        \Pi_-(mc,\mathbf{0})
        =
        \begin{pmatrix}
        0 & 0\\
        0 & 1_2
        \end{pmatrix}.
$$
Crucially note that this rank-two structure also carries the dimensionless algebra of spatial rotations. 
Defining
\begin{equation}
  \Sigma^i
  =
  \frac{i}{4}\epsilon^{ijk}
  \COM{\gamma^j}{\gamma^k},
  \label{eq:dimensionless-spin}
\end{equation}
one finds, in the standard Dirac representation, $\Sigma^i=\operatorname{diag}(\sigma^i,\sigma^i)$. 
Thus, in the rest frame, the restriction of \(\Sigma^i\) to either rank-two sector furnishes the usual two-dimensional Pauli algebra. 
For general on-shell momentum \(p_\sigma\), the corresponding internal generators are obtained by transporting the rest-frame generators with a standard boost. 
The internal algebra is therefore preserved, although its embedding in the four-dimensional Clifford representation depends on momentum.
Importantly note that the compatibility condition does not reduce $\uF_\sigma$ to a scalar as arbitrary populations and coherences within this degenerate rank-two sector commute with $\uH_{\mathrm{free}}$. 
These internal coherences should not be confused with the off-diagonal blocks $\Pi_+\uW\Pi_-$ and $\Pi_-\uW\Pi_+$ that mix the two eigenspaces.
As in the Dirac theory, $\Sigma^i$ is dimensionless and \emph{no physical angular-momentum scale has yet been introduced}.
The following identities will be useful in simplifying projected equations
\begin{equation}
    \Pi_s \gamma^\mu \Pi_s = s \frac{p^\mu}{mc}\Pi_s \text{ and }\Pi_s \gamma^\mu \Pi_{-s}= \left(\gamma^\mu + s \frac{p^\mu}{mc} \right)\Pi_{-s}
\end{equation}
The factor constraint identifies which matrix components are allowed at leading classical order. Let
\begin{equation}
    \mathsf{D}_{\uH_{\mathrm{free}}}[\mathbf{X}]
    \equiv
    \sum_{s=\pm1}\Pi_s\mathbf{X}\Pi_s
    \label{eq:block-projection}
\end{equation}
denote the operation that retains only the blocks acting within each eigenspace of $\uH_{\mathrm{free}}$ of some arbitrary $\mathbf{X}$ 
(i.e. the $\mathsf{D}$iagonal blocks relative to the eigenspaces of $\uH_{\mathrm{free}}$). These are precisely the blocks that commute with $\uH_{\mathrm{free}}$. 
The constraint and transport equations can then be separated as
\begin{equation}
    \COM{\uH_{\mathrm{free}}}{\uW_0}=\Zd,
    \qquad
    \mathsf{D}_{\uH_{\mathrm{free}}}\!\left[
        \mathcal{L}^{\mathcal{W}}_{\uH_{\mathrm{free}}}(\uW_0)
    \right]
    =\Zd.
    \label{eq:projected-classical-system}
\end{equation}
For the zero-eigenvalue sector this reduces to
\begin{equation}
    \uH_{\mathrm{free}}\uW_0=\Zd=\uW_0\uH_{\mathrm{free}},
    \qquad
    \Pi_+\mathcal{L}^{\mathcal{W}}_{\uH_{\mathrm{free}}}(\uW_0)\Pi_+=\Zd.
    \label{eq:physical-projected-system}
\end{equation}
The first relation is kinematic implying $\uW_0$ cannot contain terms that mix the subspace selected by the factor with its complementary eigenspace. 
The second applies the Weyl-ordered Liouville equation only within the selected subspace. The off-diagonal projector components are not imposed as additional classical equations. This leads us to suggest:
\begin{ansatz}[Projected Weyl-ordered transport]
\label{ans:projected-classical-transport}
We take classical matrix transport to be governed by the
Weyl-ordered Liouvillian defined in
Eq.~\eqref{extendedbracket}. Only the part acting within the
spectral sectors of \(H_{\mathrm{free}}\) is required to vanish:
\begin{equation}
    \mathcal D_{\uH_{\mathrm{free}}}
    \left[
        \mathcal L^{\mathcal W}_{\uH_{\mathrm{free}}}(\uW_0)
    \right]
    =\mathbf \Zd .
    \label{eq:projected-classical-transport}
\end{equation}
The complementary projector components are not imposed as
additional classical transport equations.
\end{ansatz}

Using the projector identities above and the fact that the free projectors are independent of $x$, the transport equation becomes
\begin{equation}
    \Pi_+
    \mathcal{L}^{\mathcal{W}}_{\uH_{\mathrm{free}}}(\uW_0)
    \Pi_+
    =
    \frac{p_\sigma^\mu}{m}\PD{\uW_0}{x^\mu}
    =\Zd.
    \label{eq:matrix-transport}
\end{equation}
Substitution of Eq.~\eqref{eq:projected-classical-ansatz} therefore gives the full $2\times2$ transport equation
\begin{equation}
    p_\sigma^\mu\PD{\uF_\sigma}{x^\mu}=0_2.
    \label{eq:internal-transport}
\end{equation}
Writing $x^0=ct$ gives
\begin{equation}
    \PD{\uF_\sigma}{t}
    +
    \sigma\frac{c^2p^i}{E_p}\PD{\uF_\sigma}{x^i}
    =0_2,
    \label{eq:coordinate-time-transport}
\end{equation}
which transports arbitrary populations and internal coherences with the correct relativistic branch velocity. It is also a phase-space continuity equation: defining the current-density $\mathbf{J}_\sigma^\mu=(p_\sigma^\mu/m)\uF_\sigma$ gives $\partial_\mu\mathbf{J}_\sigma^\mu=0_2$. 
The same conservation law holds for the trace and for every fixed matrix component.

The construction is covariant because $S(\Lambda)\Pi_+(p)S^{-1}(\Lambda)=\Pi_+(\Lambda p)$, so both the projected ansatz and Eq.~\eqref{eq:internal-transport} retain their form under a Lorentz transformation. 
The initial data on a spacelike hypersurface may therefore be any sufficiently regular $2\times2$ matrix distribution $\uF_\sigma$. 
Hermiticity and positive semidefiniteness may be imposed when $\uF_\sigma$ is interpreted as a statistical density.
Here Hermiticity and positivity refer to the statistical density within the selected two-dimensional sector, using its invariant positive inner product (the Dirac inner product restricted to that sector, with its sign chosen according to the mass-shell sheet). 
As in the covariant Dirac–Wigner formulation, these conditions should not be confused with ordinary Hermiticity or pointwise positivity of the full covariant matrix distribution.
Equation~\eqref{eq:internal-transport} transports every matrix element without any additional directional constraint. The remaining projector components acquire a different and precise role in the deformation expansion below.

If we impose that the full matrix Liouvillian vanishes, then we find that its off-diagonal blocks impose additional differential constraints on $\uW_0$ that are not required by relativistic transport and restrict otherwise admissible internal populations and coherences~\cite{everitt2026massshellfactorisationspinattempt}.
For this reason we must allow that the full matrix Liouvillian need not vanish which means that it can contain components connecting the subspace selected by the factor constraint with its complementary eigenspace. 
These components will not describe transport within the admissible classical sector, rather, they may be regarded as the matrix analogue of the reaction associated with a classical constraint.
We could therefore write the constrained Liouville equation as
\begin{equation}
    \mathcal{L}^{\uW}_{\uH_{\mathrm{free}}}(\uW_0)
    =\mathcal{R}_0, \text{ where } 
    \mathsf D_{\uH_{\mathrm{free}}}[\mathcal{R}_0]=\Zd,
\end{equation}
so $\mathcal{R}_0$ contains only components connecting distinct eigenspaces of $\uH_{\mathrm{free}}$. 
Projecting this equation onto the diagonal blocks eliminates the constraint reaction and, once more, gives
\begin{equation}
    \mathsf D_{\uH_{\mathrm{free}}}
    \left[
   		\mathcal{L}^{\uW}_{\uH_{\mathrm{free}}}(\uW_0)
    \right]=\Zd.
\end{equation}
In ordinary constrained mechanics, the reaction required to maintain a constraint is supplied by a Lagrange multiplier. 
Here the constraint acts in an internal matrix space, and every off-diagonal component connecting distinct eigenspaces of $\uH_{\mathrm{free}}$ can be represented by a commutator with it. 
The corresponding reaction therefore has the form of an internal constraint torque rather than a force in spacetime. 
We will next find that the continuation of the constraint-preservation hierarchy identifies $\mathcal{R}_0=\ui[\uH_{\mathrm{free}},\uW_1]$, so this constraint reaction, required to prevent the system from leaving the allowed constraint subspace, is supplied by the first inter-sector correction rather than by an additional classical degree of freedom.

The free generator constrained Liouville equation suggests the following pattern: the projected part at each order in $\hbar$ supplies the compatibility condition for transport within the allowed sector, while the complementary part determines the inter-sector correction at the next order. 
This structure of hierarchy of constraint-preservation conditions looks to be similar in form to the star product and commutator terms of deformation quantisation.
A complete derivation would require this hierarchy to close consistently for arbitrary phase-space generators, rather than for $\uH_{\mathrm{free}}$ alone. 
The present free generator cannot by itself determine the full deformation, since its linearity in momentum makes all higher bidifferential terms in its star commutator vanish. 
Consequently, the present calculation does not establish that the Moyal product follows from the constraint hierarchy. 
In the next section we therefore revisit this argument but now start with generators or a more general form.
Within the deformation class specified in Part II, completing the constraint-preservation hierarchy selects the Weyl–Moyal composition law and, with the additional spectral condition, the Dirac–Wigner equations.

\section{Part II -- The argument for deformation quantisation}
\subsection{Outline argument}
Part I showed that first-order mass-shell factorisation selects a rank-two spectral sector and that relativistic transport must be imposed within that sector. 
The complementary projector blocks of the matrix Liouvillian are not additional classical transport equations. 
Instead, they provide the reaction required to prevent the distribution from leaving the subspace selected by the factor constraint. 
At first order this reaction can be represented by a commutator with a correction $\uW_1$, suggesting that the projected classical equation is the first member of a more general constraint-preservation hierarchy.

The free generator $\uH_{\mathrm{free}}$ cannot determine the higher members of this hierarchy because it is linear in momentum, causing the higher bidifferential terms in its star commutator to vanish. 
We therefore make the minimal generalisation needed to expose the full structure: the four-momentum is replaced by an arbitrary commuting classical covector $\kappa^\mu(x,p)$, while the Clifford factorisation and its two-sector spectral decomposition are retained. 

The purpose of this part is to determine the deformed composition law required for the hierarchy to continue consistently to all orders. 
We begin with a general universal, local and unital bidifferential deformation of the matrix-valued phase-space algebra, without assuming the Moyal product in advance. 
Translation invariance and affine canonical covariance, together with constant-matrix neutrality, the absence of additional background structure and order-by-order homogeneity, restrict every term to a numerical coefficient multiplying a power of the canonical Poisson operator. 
Associativity of the star product then fixes these coefficients to form an exponential, while the first-order preservation condition fixes its leading coefficient. 
The resulting product is the matrix Weyl–Moyal product. Dimensional consistency further requires its deformation parameter to have units of action and comparison with established quantum mechanics identifies this parameter with $\hbar$.

Requiring constraint preservation at every order gives the vanishing matrix star commutator. 
Requiring, in addition, that the deformed distribution remain within the selected star-spectral sector produces the corresponding left and right stargenvalue equations. 
For the relativistic mass-shell factor these reproduce the Dirac–Wigner equations. 
Within the stated assumptions, the Weyl–Moyal product, quantum transport equation and spectral constraints therefore arise as successive stages in the completion of the classical constraint-preservation argument.

\subsection{The constraint preservation hierarchy}
As stated above we begin by imposing the the first assumption which will later enable us to apply translation invariance, use a constant Poisson tensor and  restrict deformation to the Poisson tensor alone is that of flat phase spacetime (Ansatz~\ref{ans:flat-phase-spacetime}). 
We note that as curved spacetime still has a canonical symplectic structure on its cotangent bundle it is possible to make the Poisson tensor constant locally it may well be possible to extend the the derivation presented here at least locally. This means that it may be possible to relax this Ansatz to include curved backgrounds where the resulting product need only be locally equivalent to the Moyal product but the global product would need to contain connection- and curvature-dependent corrections.

We next chose a more general phase space generator than that needed for the free particle which allows us to define the full hierarchy of constraint-preservation conditions. While it is not the most general form of generator possible within the expanded Clifford algebra structure it is sufficient for our needs
\begin{ansatz}[Phase space generator]\label{ans:phase-space-generator}
 Let
$ \kappa_\mu(x,p)$
 be an arbitrary classical covector of dimensions of energy. Define the two factors
\begin{equation}
 \uH_\kappa=\gamma^\mu\kappa_\mu(x,p)-mc^2 \Id,
 \ \ \ 
 \widetilde{\uH}_\kappa=\gamma^\mu\kappa_\mu(x,p)+mc^2 \Id.
 \label{eq:two-factors}
\end{equation}
The components of $\kappa_\mu(x,p)$ are ordinary commuting smooth functions. 
\end{ansatz}
While it would be interesting to explore the implications of using more general generators in classical physics that will be a subject for future work.
The Clifford relation therefore gives
\begin{equation}
 \uH_\kappa\widetilde{\uH}_\kappa
 =\left(\kappa^\mu\kappa_\mu-m^2c^4\right)\Id.
 \label{eq:factorisation}
\end{equation}
Every member of the family consequently factorises a scalar relativistic mass-shell constraint. 
We emphasise that no operator substitution and no scale of action has been introduced.

\begin{ansatz}[Timelike sector]\label{ans:timelike-sector}
We impose the restriction $\kappa^2>0$ which selects the timelike sector appropriate to a massive particle. 
In this sector there exists a local inertial frame in
which the effective four-momentum has no spatial component,
$\kappa^\mu=(\pm\Delta,\bm 0)$, where
$\Delta=\sqrt{\kappa^\mu\kappa_\mu}$.
\end{ansatz}
On the massive shell, $\Delta=mc^2$, so the two spectral sectors of $\uH_\kappa$ are separated by the nonzero energy gap $2mc^2$. 
We are therefore working in a neighbourhood of the massive shell in which these sectors remain physically distinguishable. 
Note that at $\kappa^2=0$ the rest frame and the spectral gap disappear, so the massless case requires a separate
construction rather than being included in our present argument.
\begin{ansatz}[Massive particles]
By selecting the timelike sector we assume $m > 0$. 
This is not really a new ansatz rather a corollary to the previous one. It is however interchangeable with the that ansatz as an assumption so we have listed it explicitly.  
\end{ansatz}

Following the arguments of the previous section, let us define the projectors
\begin{equation}
 \Pi_s=\frac12\left(
 \Id+s\frac{\gamma^\mu\kappa_\mu}{\Delta}
 \right),
 \qquad s=\pm1.
 \label{eq:projectors}
\end{equation}
The projectors have rank two and
\begin{equation}
 \uH_\kappa
 =(\Delta-mc^2)\Pi_+
 +(-\Delta-mc^2)\Pi_-.
 \label{eq:two-sector-spectrum}
\end{equation}
Thus there are exactly two spectral terms. 
On the shell $\Delta=mc^2$ the selected sector is the kernel of $\uH_\kappa$, and an admissible leading distribution (just as in the free particle case) has
\begin{equation}
 \uW_0=\Pi_+\uW_0\Pi_+,
 \qquad
 \uH_\kappa\uW_0=\Zd=\uW_0\uH_\kappa
 \label{eq:leading-constraint}
\end{equation}
evaluated on the on-shell where $\kappa^2=m^2c^4$.
For the momentum derivatives below, we represent the on-shell matrix density on the full $(x,p)$ space by
$$
\uW_0=\delta(q)\,\Pi_+\widetilde{\uF}\Pi_+,\qquad q=\kappa^\mu\kappa_\mu-m^2c^4,
$$
where $\ud q\ne0$ on the shell and $\widetilde{\uF}$ is any smooth extension of the on-shell density. Such extensions define the same distribution and derivatives so throughout the later discussion of star products of smooth symbols with distributional coefficients are understood term by term in the formal expansion.

As before, we can separate the spectral diagonal and complementary blocks:
\begin{align}
 \mathsf D_\kappa[\uX]
 &=\Pi_+\uX\Pi_++\Pi_-\uX\Pi_-,
 \nonumber\\
 \mathsf Q_\kappa[\uX]
 &=\Pi_+\uX\Pi_-+\Pi_-\uX\Pi_+.
 \label{eq:DQ}
\end{align}
The commutator with $\uH_\kappa$ vanishes on the first part and is invertible on the second. We now define
\begin{equation}
 \mathsf R_\kappa[\uX]
 =\frac{1}{2\Delta}
 \left(\Pi_+\uX\Pi_--\Pi_-\uX\Pi_+\right)
 \label{eq:R}
\end{equation}
The maps $\mathsf D_\kappa$, $\mathsf Q_\kappa$, and
$\mathsf R_\kappa$ are the two-sector forms of the
block-diagonal, block-off-diagonal, and energy-denominator
superoperators used in Schrieffer--Wolff perturbation theory
\cite{BRAVYI20112793}. 
Now direct multiplication gives
\begin{equation}
 \mathsf D_\kappa\!\left[\COM{\uH_\kappa}{\uX}\right]=\Zd,
 \qquad
 \COM{\uH_\kappa}{\mathsf R_\kappa[\uX]}
 =\mathsf Q_\kappa[\uX].
 \label{eq:spectral-identities}
\end{equation}
Let us also define 
\newcommand{\cP}{{\mathcal P}}
\begin{equation}
\cP = \overleftarrow{\PD{}{p_\mu}}\,\overrightarrow{\PD{}{x^\mu}}
 -\overleftarrow{\PD{}{x^\mu}}\,\overrightarrow{\PD{}{p_\mu}}. 
\end{equation}
For matrix-valued phase-space functions this defines
\begin{equation}
 \uF\mathcal P\uG
 =\PD{\uF}{p_\mu}\PD{\uG}{x^\mu}
 -\PD{\uF}{x^\mu}\PD{\uG}{p_\mu}.
 \label{eq:P}
\end{equation}
The derivatives act componentwise and the displayed matrix order is retained. The Weyl-ordered classical Liouvillian is then
\begin{equation}
 \mathcal L^{\mathcal W}_{\uF}(\uG)
 =\frac12\left(\uF\mathcal P\uG-\uG\mathcal P\uF\right).
 \label{eq:Weyl-L}
\end{equation}
For commuting scalar functions this is the usual Poisson bracket. Just as for the free particle, for matrices it is an antisymmetric bilinear differential expression rather than a Lie bracket.
The use of this ordering earlier was provisional: its appearance as the derivative term of a star commutator will follow from the deformation conditions below.

\begin{ansatz}[The classical preservation condition]
\label{az:classical-preservation}Following the free-particle result, we require the Weyl-ordered Liouvillian of the on-shell matrix distribution $\uW_0$ to vanish in the spectral-diagonal blocks:
\begin{equation}
 \mathsf D_\kappa\!\left[
 \mathcal L^{\mathcal W}_{\uH_\kappa}(\uW_0)
 \right]=\Zd.
 \label{eq:first-preservation}
\end{equation}
Note that while  we have explicitly required the Weyl-ordered Liouvillian here, later ansatz such as translation invariance and linear symplectic covariance will combine to add sufficient constraints that this ordering is necessary, hence making the specification here motivational and logically redundant.
\end{ansatz}
By Eqs.~\eqref{eq:DQ}--\eqref{eq:spectral-identities}, this is equivalent to the existence of a complementary matrix $\uW_1$ satisfying
\begin{equation}
 \COM{\uH_\kappa}{\uW_1}
 +\ui\mathcal L^{\mathcal W}_{\uH_\kappa}(\uW_0)=\Zd.
 \label{eq:first-order}
\end{equation}
Equations~\eqref{eq:first-preservation} and \eqref{eq:first-order} have the same algebraic form as the first-order block-decoupling equation in Schrieffer--Wolff perturbation theory~\cite{BRAVYI20112793}. 
There an off-diagonal perturbation is cancelled by a commutator with the first-order generator, by contrast, here the off-diagonal part of the classical Liouvillian is cancelled by the commutator $\COM{\uH_\kappa}{\uW_1}$. 
The algebraic mechanism is therefore the same, although $\uW_1$ is an inter-sector correction to the ensemble distribution rather than the generator of a unitary
transformation.
Note fixing the factor $\ui$ in the above equation is a phase convention.
Indeed, if Eq.~\eqref{eq:first-preservation} holds, one may take
\begin{equation}
 \mathsf Q_\kappa[\uW_1]
 =-\ui\mathsf R_\kappa\!\left[
 \mathcal L^{\mathcal W}_{\uH_\kappa}(\uW_0)
 \right]
 \label{eq:first-reaction}
\end{equation}
where $\mathsf D_\kappa[\uW_1]=0$.
Conversely, applying $\mathsf D_\kappa$ to Eq.~\eqref{eq:first-order} gives Eq.~\eqref{eq:first-preservation}. The first correction therefore supplies the reaction needed to preserve the classical spectral constraint.

To determine whether this first reaction can be continued
consistently, introduce a formal real constant $\lambda$ and regard $W_1$ as the first correction
to the leading distribution:
\begin{equation} \label{eq:Wlambda}
    \uW_{\lambda}
    = \uW_0+\lambda \uW_1+\Order{\lambda^2},
\end{equation}
so that $\uW_{\lambda}\big|_{\lambda=0}=\uW_0 $. At this stage, $\lambda$ is not identified with Planck's constant and will be set only when comparison to existing theories is possible (although we will later show that it must have units of action). We also wait to introduce a specific ansatz for $\lambda$ until a later  point in the argument where justification of its units will arise naturally.

\subsection{The star product}

Because the leading distribution $\uW_0$ lies in the kernel of $\uH_\kappa$ on the mass shell, Eq.~\eqref{eq:leading-constraint} gives $\COM{\uH_\kappa}{\uW_0}=\Zd$ at zeroth order, while the classical preservation condition of Ansatz~\ref{az:classical-preservation} leads to the first-order reaction Eq.~\eqref{eq:first-order}.
By Weyl-ordered classical Liouvillian Eq.~\eqref{eq:Weyl-L}, this equation contains an ordinary matrix commutator and a derivative contribution proportional to $\uH_\kappa\mathcal P\uW_0-\uW_0\mathcal P\uH_\kappa$.
Both orders thus have the same antisymmetrised left--right
structure. This suggests an explicit requirement for their
continuation:
\begin{ansatz}[Self-preservation of a generator]
We require the all-order preservation rule to be bilinear on
matrix-valued phase-space functions and to vanish when the same
function is used as both generator and argument.
\end{ansatz}
Applying this requirement to the sum of any two functions shows that exchanging the generator and argument must reverse the sign of the rule at every order. 
We therefore seek a deformation of ordinary multiplication, denoted by $\star$, whose commutator reproduces the known zeroth- and first-order conditions and provides the higher-order terms.

We start by imposing a universality requirement. 
Note that as every $\uH_\kappa$ is a generator on the same canonical phase space, $\kappa_\mu(x,p)$ specifies the dynamics, whereas the rule for composing phase-space quantities belongs to the kinematic structure of that phase space. 
Just as the classical pointwise product and Poisson bracket do not depend on the Hamiltonian or external field, their deformation should be a single product acting on the entire phase-space algebra. 
\begin{ansatz}[Universality]\label{ans:universality}
We assume that there is nothing special about $\uH_{\kappa}$  and can define a star product in terms of functions $\uF$ and $\uG$ alone.
\end{ansatz}

We therefore seek to define one product, independent of the chosen $\kappa_\mu(x,p)$ as follows:

\begin{ansatz}[Star product] The rule for composing phase-space quantities is defined by a product of the form
\begin{equation}
 \uF\star\uG
 =\uF\uG+\sum_{n=1}^{\infty}\lambda^n C_n(\uF,\uG).
 \label{eq:general-star}
\end{equation}
where $C_n$ is a bilinear finite-order bidifferential operator on the formal matrix-valued phase-space
algebra, that $C_0(\uF,\uG)=\uF\uG$ where the product is local, unital/normalised and universal. 
\end{ansatz}
We also define the star commutator
\begin{align}
 \MB{\uF}{\uG} 
 \equiv
 \uF\star \uG
 -\uG\star \uF
\end{align}

Substitution of $\uH_{\kappa}$ and $\uW_{\lambda}$ into this  expansion and star commutator gives
\begin{align}
 &\MB{\uH_{\kappa}}{\uW_{\lambda}} 
  =
 \COM{\uH_{\kappa}}{\uW_0} 
 \\ &\ \ \ \ 
 +\lambda \,
 \Big[
 	\COM{\uH_{\kappa}}{W_1}
 	+
	C_1(\uH_{\kappa},\uW_0)
 	-
	C_1(\uW_0,\uH_{\kappa})
 \Big]
 +\Order{\lambda^2}. \nonumber
\end{align}
The zeroth-order term vanishes by Eq.~\eqref{eq:leading-constraint}. 
Requiring the coefficient of $\lambda$    to reproduce Eq.~\eqref{eq:first-order} therefore gives
\begin{equation}
	 C_1(\uF,\uG)-C_1(\uG,\uF)
 	=
	\ui \mathcal{L}^{\uW}_{\uF}(\uG),\label{eq:ucc}
\end{equation}
where the we have used universality ansatz above to realise correspondence of the first order term when the particular pair $\uH_{\kappa}$ and $\uW_0$ are used. Again, the factor $\ui$ is the chosen reality/phase convention for real $\lambda$.

This determines only the part of $C_1$ that is antisymmetric under interchange of its arguments. 
A candidate consistent with Eq.~\eqref{eq:ucc} is
\begin{equation}
    C_1(\uF,\uG)=\frac{\ui}{2}\uF\cP \uG.
    \label{eq:C1}
\end{equation}
Equation~\eqref{eq:C1} is introduced provisionally to connect the product to the preservation hierarchy. 
Its form will be obtained below from the stated symmetry and homogeneity assumptions, so no separate first-order Weyl ansatz is needed in the final argument.
Equation~\eqref{eq:ucc} fixes only the antisymmetric part of $C_1$ and the provisional form in Eq.~\eqref{eq:C1} has not been deduced from preservation alone. 
Translation invariance, linear symplectic covariance, constant-matrix neutrality, absence of further background tensors and order-by-order homogeneity will imply Eq.~\eqref{eq:Cn-natural-form}. 
At $n=1$ this is $C_1=a_1\uF\mathcal P\uG$, and Eq.~\eqref{eq:ucc} then gives $a_1=\ui/2$. 
Thus the first-order Weyl form, and the matrix Liouvillian anticipated in Part I, are selected within the stated deformation class.
Possible symmetric ordering terms outside this restricted class are not analysed here and we note that their inclusion would alter stargenvalue equations even when the first-order star commutator is unchanged.

\begin{ansatz}[Symplectic form] \label{az:symplectic} We impose invariance under  linear symplectic covariance as this is known to preserve the fundamental structure of physical laws and phase space geometry in classical and quantum mechanics 
(in particular this preserves the canonical Poisson bracket)
\end{ansatz}

\begin{ansatz}[Translation invariance of $\star$] Flat phase space has no preferred origin. We therefore require the product to be invariant under translations of both $x^\mu$ and $p_\mu$.
\end{ansatz}
These are both symmetries of the classical canonical Poisson bracket on flat phase space: its coefficients are constant, and it is covariant under canonical transformations, including the linear symplectic maps used here. 
This pair of assumptions thus concerns the bracket as a kinematic rule, not the particular generator $\uH_\kappa$ or a chosen state as external fields, for example, may break translation symmetry of the dynamics. 
Requiring the star product itself to retain translation and linear symplectic covariance extends these classical properties to the subsequently derived deformation.

The product consequently uses only the canonical phase-space structure represented by $\cP$ so that it introduces no preferred position, momentum, direction, or additional dynamical tensor. Related arguments can be found in~\cite{Howe1989,gutt_natural_2003,BAYEN197861}. We will return to this in more detail in Ansatz~\ref{az:bg}.

The bidifferential character of the deformation makes it natural to leave the internal matrix algebra undeformed.
That is for any two constant matrices $\uA$ and $\uB$ that $ \uA \star \uB = \uA\uB$. We go further and also impose:
\begin{ansatz}[Constant matrix neutrality] Given that differentials of a constant matrix will vanish we can take this further and consider products of any constant $4\times 4$ matrix $\uA$ with an arbitrary matrix-valued phase-space function $\uF$ we can thus assert that 
\begin{equation}
  \uA \star \uF =\uA \uF
  \text{ and }
  \uF \star \uA = \uF \uA,
  \label{eq:AstarF}
\end{equation}
\end{ansatz}
This Ansatz excludes terms of differential order zero in only one argument and acts as a normalisation.
For scalar symbols the corresponding property follows from complex bilinearity and unitality, since every constant is a multiple of the identity. A constant matrix need not be, so requiring it to multiply an arbitrary matrix-valued symbol in the ordinary way on both sides is an additional condition.
We now impose associativity of the star product as this ensures that the algebraic action of the star commutator
$\MB{F}{G}$ obeys the product rule and the Jacobi identity.
These properties ensure the consistent closure of generators and constraints. Associativity ensures that extending the hierarchy to further products or constraints does not introduce bracket-dependent results (and it would not be possible to proceed further without some other constraint to determine them).
\begin{ansatz}[Associativity of $\star$]\label{ans:associativity}
We require the star product to be associative over arbitrary $4\times 4$ matrix-valued phase-space functions
$$
  (\uF \star \uG)\star \uK = \uF \star (\uG \star \uK).
$$
Associativity ensures the star commutator will satisfy Jacobi and the Leibniz rule with respect to the star product.
\end{ansatz} 
\noindent Together with Eq.~\eqref{eq:AstarF}, associativity 
gives for a constant matrix $\uA$
\begin{align}
  	(\uA \uF)\star \uG 
  	&= 
	\uA( \uF \star \uG), 
	\label{eq:left}\\
	(\uF \uA) \star \uG 
	&= 
	\uF \star(\uA \uG),
	\label{eq:middle}\\
	\uF \star (\uG \uA) 
	&= 
	(\uF\star \uG) \uA.                                      \label{eq:right}
\end{align}
Now we substitute 
$$
  \uF \star \uG
  =\sum_{n=0}^\infty \lambda^n C_n(\uF,\uG)
$$
into Eqs.~\eqref{eq:left}--\eqref{eq:right} and compare equal powers of $\lambda$.  
For every constant matrix $A$ we obtain
\begin{align}
  C_n(\uA \uF,\uG) 
  &= 
  \uA\,C_n(\uF,\uG),
  \label{eq:Cn-left}\\ 
  C_n(\uF \uA, \uG) 
  &= 
  C_n(\uF,\uA \uG),
  \label{eq:Cn-mid}\\
  C_n(\uF,\uG\uA) 
  &= 
  C_n(\uF,\uG)\,\uA.                                         \label{eq:Cn-right}
\end{align}
Thus the deformation acts only on the phase-space dependence and never changes the order of the matrix factors.
We now use a commutant argument for full matrix algebras,
followed by expansion in matrix units~\cite{EtingofEtAl2011}.
First we restrict $C_n$ to arguments proportional to the identity. 
For scalar phase-space functions $f(x,p)$ and $g(x,p)$ we define the matrix-valued function
\begin{equation}
  X_n(f,g):=C_n(f\Id,g\Id).
\end{equation}
For any constant $\uA$, Eqs.~\eqref{eq:Cn-left}--\eqref{eq:Cn-right} give
\begin{align}
  	\uA X_n(f,g)
   	&=
	C_n(\uA f\Id, g\Id) \nonumber\\
   	&=
	C_n(f\Id,\uA g\Id) \nonumber\\
   	&=
	X_n(f,g)\uA.                                                      \label{eq:Xn-central}
\end{align}
Thus $X_n(f,g)$ commutes with every constant $4\times 4$ matrix.  
As only $4\times4$ matrices that commute with every other $4\times4$ matrix are scalar multiples of the identity there must exist a scalar bidifferential operator $c_n$ such that
\begin{equation}
  C_n(f\Id,g\Id)
     =c_n(f,g)\Id.                                          \label{eq:scalar-Cn}
\end{equation}
Now completely general matrix-valued functions can be expanded in terms of the standard matrix units
\[
  \uF=\sum_{a,\,b}\uF_{ab}\uE_{ab},
  \qquad
  \uG=\sum_{c,\,d}\uG_{cd}\uE_{cd},
\]
where $\uE_{ab}$ and $(\uE_{ab})_{ij}=\delta_{ai}\delta_{bj}$, so that $\uE_{ab}\uE_{cd}=\delta_{bc}\uE_{ad}$.  
Since $f\uE_{ab}=\uE_{ab}(f\Id)$ and $g \uE_{cd}=(g\Id)\uE_{cd}$, Eqs.~\eqref{eq:Cn-left} and \eqref{eq:Cn-right} imply
\begin{align}
  	C_n(f\uE_{ab},g\uE_{cd})
   	&=
	\uE_{ab}\,C_n(f\Id,g\Id)\,\uE_{cd} 
	\nonumber\\
   	&
	=c_n(f,g)\uE_{ab} \uE_{cd} 
   	\nonumber\\
   	&=
	\delta_{bc}\,c_n(f,g) \uE_{ad}.                                   \end{align}
which together with bilinearity gives
\begin{align}
  	C_n(\uF,\uG)
    &=
    \sum_{a,b,c,d}
    	C_n(
			\uF_{ab} \uE_{ab},
			\uG_{cd} \uE_{cd}
			)
	\nonumber\\
    &=
    \sum_{a,b,d}
    	c_n(
			\uF_{ab},\uG_{bd}
			)\,
		\uE_{ad},
\end{align}
or, component by component,
\begin{equation}
  	\left[C_n(\uF,\uG)\right]_{ad}
    =
    \sum_b 
    	c_n(\uF_{ab},\uG_{bd}).                                    \label{eq:Cn-component}
\end{equation}
The repeated index $b$ is therefore contracted exactly as in ordinary matrix multiplication.  To determine $c_n$ we now impose translation invariance as this makes the coefficients of $c_n$ constant.  
A general term can therefore be written
\begin{equation}
  T^{\alpha_1\ldots\alpha_r\,
     \beta_1\ldots\beta_s}
  (\partial_{\alpha_1}\cdots\partial_{\alpha_r}f)
  (\partial_{\beta_1}\cdots\partial_{\beta_s}g).                    \label{eq:general-scalar-bidifferential}
\end{equation}
Affine canonical covariance requires the coefficient tensor $T$ to be symplectically invariant.  
Under the assumption that no additional phase-space tensor is available, such a coefficient is constructed by contracting derivative indices in pairs with the Poisson tensor $P^{\alpha\beta}$ (see, for example, discussion on first fundamental theorem of invariant theory for the symplectic group in~\cite{GoodmanWallach2009}).  
\begin{ansatz}[No additional background structure]\label{az:bg}
We assume that the deformation is constructed only from the canonical structure of phase space and the formal parameter $\lambda$ already appearing in the expansion. 
In particular, no further fixed phase-space tensors or independent constants with the dimensions of action are introduced. The coefficients ($C_n$) must therefore be built only from derivatives, the canonical Poisson tensor ($P^{\alpha\beta}$), and dimensionless numerical constants.
\end{ansatz}
The ordinary canonical Poisson bracket uses no phase-space tensor beyond $P^{\alpha\beta}$, but extending this restriction to every $C_n$ is an additional assumption. It constrains the universal composition rule, not $\uH_\kappa$, which may contain a spacetime metric or externally specified fields.
To define the Poisson tensor, let $z\,^\alpha=(x\,^\mu,p_\mu) $
denote the canonical coordinates on the one-particle phase space.  
The canonical Poisson tensor is then the constant antisymmetric tensor $ P^{\alpha\beta}\equiv\{z\,^\alpha,z\,^\beta\}$ so that for any scalar functions $f$ and $g$,
\begin{align}
  f\mathcal{P}g
   =
     P^{\alpha\beta}
     (\partial_\alpha f)(\partial_\beta g)
   =
     \PD {f}{p_\mu}
     \PD{g}{x^\mu}
     -
     \PD{f}{x^\mu}
     \PD{g}{p_\mu}.                            
\end{align}
Thus $P^{\alpha\beta}$ denotes the Poisson tensor itself, whereas $\mathcal{P}$ between two functions denotes the corresponding bidifferential operation.
For matrix-valued functions the derivatives act entry by entry and the resulting matrices are multiplied in their displayed order.
Homogeneity of degree $n$ means that exactly $n$ copies of $P^{\alpha\beta}$ occur in $c_n$. Specifically:
\begin{ansatz}[Order-by-order homogeneity]\label{ans:order-homogeneity} 
We assume that the coefficient $C_n$ multiplying $\lambda^n$ contains exactly $n$ copies of the canonical Poisson tensor $P^{\alpha\beta}$. 
Equivalently, $C_n(\uF,\uG)$ is of differential order $n$ in each argument: each additional power of $\lambda$ introduces one additional Poisson contraction and one additional derivative acting on both $f$ and $g$. 
\end{ansatz}
We note there is no order-by-order homogeneity requirement in ordinary Liouville dynamics, which has no deformation series. It is a further selection criterion for the all-order product.
Without this restriction, associativity would also permit the more general form $\uF \star \uG = \uF \exp[f(\lambda)\mathcal P]\uG$, where $f(\lambda)=\ui\lambda/2+O(\lambda^2)$, rather than, as we shall later see, selecting the Moyal product of the standard form.

Now note that no copy of $P^{\alpha\beta}$ can contract two derivatives acting on the same function.  
Indeed, commuting partial derivatives are symmetric whereas
the Poisson tensor is antisymmetric, and hence
\begin{equation}
  P^{\alpha\beta}\partial_\alpha\partial_\beta\, f=0,
  \qquad
  P^{\alpha\beta}\partial_\alpha\partial_\beta\, g=0.                 \label{eq:same-factor-contraction}
\end{equation}
The same cancellation occurs if further derivatives are present.  Every
copy of \(P^{\alpha\beta}\) must therefore connect one derivative of \(f\)
to one derivative of \(g\).  It follows that \(r=s=n\).

The most general remaining contraction is a linear combination of the possible pairings,
\begin{align}
  c_n(f,g)
   =\sum_{\sigma\in S_n}b_\sigma\,
     &P^{\alpha_1\beta_{\sigma(1)}}\cdots
      P^{\alpha_n\beta_{\sigma(n)}} \nonumber\\[-0.5ex]
     &\times
      (\partial_{\alpha_1}\cdots\partial_{\alpha_n}f)
      (\partial_{\beta_1}\cdots\partial_{\beta_n}g).                \label{eq:all-symplectic-pairings}
\end{align}
where $S_n$ denotes the symmetric group of all permutations of the numbers $1 \ldots n$ and each $\sigma$ is one particular permutation. 
Both derivative tensors are symmetric in their respective indices, so all permutations $\sigma$ must give the same operator after relabelling dummy indices.  
Their coefficients therefore combine into the single number
$a_n \equiv \sum_{\sigma\in S_n}b_\sigma$.  
Consequently
\begin{equation}
  c_n(f,g)
   =
   a_n P^{\alpha_1\beta_1}\cdots P^{\alpha_n\beta_n}
      (\partial_{\alpha_1}\cdots\partial_{\alpha_n}f)
      (\partial_{\beta_1}\cdots\partial_{\beta_n}g)
   =
   a_n f \mathcal{P}^ng.                                                       \label{eq:cn-scalar-form}
\end{equation}
and substitution into Eq.~\eqref{eq:Cn-component} yields
\begin{equation}
  \left[C_n(\uF,\uG)\right]_{ad}
     =a_n\sum_b \uF_{ab} \mathcal P^n \uG_{bd}
     =\left[a_n\uF \mathcal P^n \uG\right]_{ad}.                                    \label{eq:Cn-final-component}
\end{equation}
Hence, in matrix notation,
\begin{equation}
  C_n(\uF,\uG)=a_n\,\uF\, \mathcal{P}^n \uG,
  \text{ with } a_0=1,                                                     \label{eq:Cn-natural-form}
\end{equation}
where the derivatives in $P^n$ act entry by entry and the differentiated
matrices are multiplied in the displayed order. The value $a_0=1$ is fixed by the undeformed zeroth-order product $C_0(\uF,\uG)=\uF \uG$.

It now remains to determine the constants $a_n$.  
To avoid confusing the generating series with the scalar phase-space functions used above, we will denote it by
\begin{equation}
  \Phi(t)=\sum_{n\geq0}a_n t^n,
  \qquad
  \Phi(0)=a_0=1.                                                    \label{eq:Phi-definition}
\end{equation}
Our current argument is restricted to one physical phase space, with coordinates $(x,p)$ (although extension would be straight forward).  
To indicate which matrix-valued factor a derivative acts upon, we now define
\begin{equation}
  P_{\uF \uG}
   :=
   \frac{\partial^{(\uF)}}{\partial p_\mu}
   \frac{\partial^{(\uG)}}{\partial x^\mu}
   -
   \frac{\partial^{(\uF)}}{\partial x^\mu}
   \frac{\partial^{(\uG)}}{\partial p_\mu}. 
   \label{eq:PFG}                          
\end{equation}
The superscript labels specify the factor to be differentiated and do not introduce additional coordinates or particles.  
Repeated phase-space indices are summed, and, in particular, $P_{\uF \uG}(\uF \uG)=\uF \mathcal{P} \uG$.  

Equation \eqref{eq:general-star} can now be written, using Eqns.~\eqref{eq:Phi-definition} and \eqref{eq:PFG}, as
\begin{equation}
  \uF \star \uG = 
  \sum_{n=0}^\infty 
  a_n \lambda^n \, \uF\mathcal{P}^n \uG
  =
  \Phi(\lambda P_{\uF \uG})\,\uF \uG,                               \label{eq:star-labelled-derivatives}
\end{equation}
where $\uF\uG$ is ordinary matrix multiplication in the displayed order.

It is now possible to determine the \emph{dimensions of the deformation parameter} $\lambda$ as they are fixed by the structure of the above expansion. 
The product $x^\mu p_\mu$ has dimensions of action: $x^0p_0=(ct)(E/c)=Et$, while each spatial product $x^i p_i$ has dimensions of length times momentum. 
The Poisson operator $\mathcal{P}$, which contains one derivative with respect to $x^\mu$ and one with respect to $p_\mu$, therefore has dimensions of inverse action. Consequently,
\begin{equation}
    [\uF \mathcal{P}^n \uG]
      =[\uF][\uG]\,[\text{action}]^{-n}.
\end{equation}
By Ansatz~\ref{az:bg} the coefficients $a_n$ are dimensionless, and every term $a_n\lambda^n \uF \mathcal{P}^n\uG$ in Eq.~\eqref{eq:star-labelled-derivatives} must have the same dimensions as the undeformed product $\uF\uG$. It follows that
\begin{equation}
    [\lambda]=[\mathrm{action}],
\end{equation}
This conclusion introduces no quantum assumption and does not yet identify $\lambda$ with
Planck's constant rather it shows only that the deformation parameter must have the same physical dimensions as $\hbar$.

Note that when $\uF\uG$ is star-multiplied by $\uK$, the derivative acting on the first argument acts on the product $\uF\uG$.  
The ordinary product rule gives
$$
  \partial(\uF \uG)
  =
  (\partial \uF)\uG+\uF(\partial \uG),
$$
so the Poisson operator coupling $\uF\uG$ to $\uK$ is
$P_{\uF\uK}+P_{\uG\uK}$.  
Similarly, the operator coupling $\uF$ to $\uG\uK$ is $P_{\uF\uG}+P_{\uF\uK}$.  
It follows that
\begin{align}
  (\uF\star \uG)\star \uK
   &=
     \Phi(\lambda P_{\uF\uG})
     \Phi\bigl(\lambda(P_{\uF\uK}+P_{\uG\uK})\bigr)\uF\uG\uK,
     \label{eq:left-associated-Phi}\\
  \uF\star(\uG\star \uK)
   &=
     \Phi(\lambda P_{\uG\uK})
     \Phi\bigl(\lambda(P_{\uF\uG}+P_{\uF\uK})\bigr)\uF\uG\uK.                     \label{eq:right-associated-Phi}
\end{align}
The matrix order is $\uF\uG\uK$ on both sides.  
Because the Poisson tensor is constant, $P_{\uF\uG}$, $P_{\uF\uK}$, and $P_{\uG\uK}$ commute.  
Associativity for arbitrary $\uF,\uG$ and $\uK$ therefore gives an equality between constant-coefficient differential operators.  
Comparing their symbols, or equivalently choosing the derivative data of $\uF,\uG$ and $\uK$ independently at a point, allows the three pairwise Poisson contractions to be treated as commuting formal variables.
Already one canonical pair is sufficient for these three contractions to range over an open set of triples as equality there implies equality of their formal power-series coefficients.  
Thus associativity requires
\begin{equation}
  \Phi(u)\Phi(v+w)=\Phi(v)\Phi(u+w).                               \label{eq:Phi-cocycle}
\end{equation}
for commuting formal variables $u,v,w$, corresponding respectively to
$\lambda P_{\uF\uG}$, $\lambda P_{\uG\uK}$, and $\lambda P_{\uF\uK}$.  
Taking $v=0$ and using $\Phi(0)=1$ yields
\begin{equation}
  \Phi(u+w)=\Phi(u)\Phi(w).                                        \label{eq:Phi-Cauchy}
\end{equation}
Differentiation with respect to $w$ at $w=0$ gives
$\Phi'(u)=a_1\Phi(u)$, and hence
\begin{equation}
  \Phi(t)=\exp(a_1\,t),
  \qquad
  a_n=\frac{a_1^n}{n!}.                                             \label{eq:an-exponential}
\end{equation}

The first-order antisymmetry condition already obtained above is
\begin{equation}
  C_1(\uF,\uG)-C_1(\uG,\uF)
  =
  \ui \mathcal L_\uF^\mathcal W(\uG)
   =
   \frac{\ui}{2}\left(\uF \mathcal{P}\uG-\uG \mathcal P \uF\right).
\end{equation}
Substituting Eq.~\eqref{eq:Cn-natural-form} at $n=1$ into the antisymmetry condition fixes $a_1=\ui/2$ and recovers the provisional Eq.~\eqref{eq:C1}. We therefore
obtain the matrix-valued Weyl--Moyal product
\begin{equation}
  \uF\star \uG
    =\uF\exp\!\left(\frac{\ui\lambda}{2}\mathcal{P}\right)\uG
    =\sum_{n\geq0}\frac{1}{n!}
       \left(\frac{\ui\lambda}{2}\right)^n \uF \mathcal{P}^n \uG .
  \label{eq:matrix-moyal-product}
\end{equation}
No commutativity of $\uF$ and $\uG$, and no common scalar phase-space factor, has been used.  
The statement is Weyl ordered in the phase-space variables, while the internal matrix factors retain their original order.  

\begin{ansatz}[Planck's constant]\label{ans:planck-constant}
By comparison with known results in non-relativistic quantum mechanics we assert $\lambda$ which has the units of action is indeed Planck's constant $\hbar$ (if we did not have this historical imperative its value would have to be determined through experiment). 
\end{ansatz}
\begin{equation}
  \boxed{
  \uF\star \uG
    =\uF\exp\!\left(\frac{\ui\hbar}{2}\mathcal{P}\right)\uG
    =\sum_{n\geq0}\frac{1}{n!}
       \left(\frac{\ui\hbar}{2}\right)^n \uF \mathcal{P}^n \uG .}
  \label{eq:matrix-moyal-product}
\end{equation}

The appearance of $\hbar$ in the deformation now supplies the physical angular-momentum scale. 
This means that in the rest frame, the corresponding spin generators are
\begin{equation}
  S^i
  =\frac{\hbar}{2}\Sigma^i,
  \qquad
  \COM{S^i}{S^j}
  =i\hbar\epsilon^{ijk}S^{\,k}.
  \label{eq:physical-spin}
\end{equation}
The Clifford factorisation therefore supplies the two-state spinor algebra, while deformation quantisation supplies its physical scale through $\lambda=\hbar$.

\subsection{Completion of the constraint-preservation hierarchy.}
We can now return to the hierarchy that motivated the deformation.
Complete the formal expansion introduced in Eq.~\eqref{eq:Wlambda} by writing
\begin{equation}
    \uW_\lambda=\sum_{n=0}^{\infty}\hbar^n \uW_n .
    \label{eq:W-formal-series}
\end{equation}
Since $\uH_\kappa$ is independent of $\hbar$ (as we are assuming for the moment only classical generators), its star
commutator with $\uW_\lambda$ can be expanded as
\begin{align}
\MB{\uH_\kappa}{\uW_\lambda}
   &=\sum_{N=0}^{\infty}\hbar^N\mathfrak E_N, \\
 \mathfrak E_N
   &=
   \sum_{r=0}^{N}
   \left[
      C_r(\uH_\kappa,\uW_{N-r})
      -C_r(\uW_{N-r},\uH_\kappa)
   \right],
 \label{eq:preservation-hierarchy}
\end{align}
where $C_0(\uF,\uG)=\uF\uG$. The first two members of this hierarchy are precisely the conditions already obtained. At zeroth order on $\hbar$,
\begin{equation}
    \mathfrak E_0=\COM{\uH_\kappa}{\uW_0}=\Zd,
\end{equation}
which is the compatibility condition in Eq.~\eqref{eq:leading-constraint}. 
At first order, Eq.~\eqref{eq:ucc} gives
\begin{equation}
 \mathfrak E_1
   =\COM{\uH_\kappa}{\uW_1}
     +i\mathcal{L}^{\mathcal W}_{\uH_\kappa}(\uW_0)=\Zd,
\end{equation}
which is exactly Eq.~\eqref{eq:first-order}.

At every higher order, the same spectral decomposition has the
same interpretation. Because
$D_\kappa[[\uH_\kappa,\uX]]=0$, the diagonal projection of
$\mathfrak E_N=0$ supplies the next compatibility condition
within the allowed sector. Its complementary projection then
determines the inter-sector part of $\uW_N$ through the inverse map $\mathsf R_\kappa$. Following this hierarchy up the chain order at a time leads to:

\begin{ansatz}[All-order constraint preservation]\label{ans:all-order-preservation}
We require an admissible deformed distribution to satisfy every member of the constraint-preservation hierarchy,
$\mathfrak E_N=0$ for all $N$. 
\end{ansatz}

When taken to all orders, this is exactly equivalently to saying
\begin{equation}
       \MB{\uH_\kappa}{\uW_\lambda}
       =
       \uH_\kappa\star \uW_\lambda-\uW_\lambda\star \uH_\kappa
       =\Zd .
       \label{eq:star-preservation}
\end{equation}
Thus the vanishing star commutator is not an additional
order-by-order formula: it is the compact all-orders form of the preservation hierarchy whose zeroth- and first-order members have already been established. 
In particular, it does not impose the unnecessarily strong classical condition $\mathcal{L}^{\uW}_{\uH_\kappa}(\uW_0)=\Zd$ as its complementary part is cancelled by the commutator with $\uW_1$, as in Eq.~\eqref{eq:first-order}.

Equation~\eqref{eq:star-preservation} implies that the left and right star products are equal, but it does not by itself imply that either is proportional to $\uW_\lambda$. 
That additional information comes from the factor constraint in Eq.~\eqref{eq:leading-constraint}, which places $\uW_0$ entirely within one spectral sector of $\uH_\kappa$.

\begin{ansatz}[Preservation of the selected spectral sector]\label{ans:star-spectral-sector}
We require the deformation of $\,\uW_0$ to remain within the corresponding star-spectral sector. 
\end{ansatz}
To satisfy this ansatz it is sufficient to impose the left star constraint
\begin{equation}
    (\uH_\kappa-\epsilon\Id)\star \uW_\lambda=0,
    \label{eq:left-star-constraint}
\end{equation}
where $\epsilon$ is a scalar stargenvalue. 
Since $\epsilon\Id$ is constant and is therefore unaffected by the star product, Eq.~\eqref{eq:star-preservation} immediately also gives the right constraint. Hence
\begin{equation}
    \uH_\kappa\star \uW_\lambda
       =\epsilon \uW_\lambda
       =\uW_\lambda\star \uH_\kappa .
    \label{eq:two-sided-stargenvalue}
\end{equation}
For the mass-shell factor selected in Eq.~\eqref{eq:leading-constraint}, the relevant sector has $\epsilon=0$, and the equations reduce to
\begin{equation}
\boxed{
    \uH_\kappa\star \uW_\lambda=0
    =\uW_\lambda\star \uH_\kappa .
    \label{eq:two-sided-zero-constraint}}
\end{equation}
This has the same form as the time-independent Schr\"odinger Wigner/Moyal stargenvalue equations of nonrelativistic quantum mechanics but in phase spacetime.


\section{Comparison to the Dirac-Wigner formalism}

Following~\cite{D1,D2,D3,D4} we seek a direct comparison of the matrix-valued Moyal equation with the Dirac equation by considering the Wigner transform of the Dirac density matrix.  Let $\psi(x)$ satisfy
the minimally coupled Dirac equation
$$
    \left[
        \ui\hbar c\gamma^\mu
        \left(
            \PD{}{x^\mu}
            +
            \frac{iq}{\hbar c}A_\mu
        \right)
        - mc^2
    \right]
    \psi(x)
    =
    0.
$$
The spinor density is then defined as
$
    \rho_{\alpha\beta}(x_1,x_2)
    \equiv
    \psi_\alpha(x_1)\bar\psi_\beta(x_2),
$
where $\bar \psi = \psi^\dag \gamma^0$ is the Dirac adjoint.
We now introduce the centre and relative coordinates
$
\xi=\frac{x_1+x_2}{2}$ and $
y=x_1-x_2.
$
A spinor-matrix Wigner function is then given by performing a Wigner transformation (i.e. a Fourier transform with respect to the coordinate separation) on $\rho_{\alpha\beta}(x_1,x_2)$, namely
$$
W_{\alpha\beta}(\xi,p)
=
\int \ud^4y\,
\exp\!\left(\frac{i}{\hbar}p_\mu y^\mu\right)
\rho_{\alpha\beta}\left(\xi+\frac{y}{2},\xi-\frac{y}{2}\right),
$$
The expression above is the ordinary fixed-gauge Wigner transform.
A fully gauge-covariant Wigner function instead inserts a Wilson line between $x_1$ and $x_2$~\cite{D1,wilsonLine}. Our treatment could therefore be regarded as a fixed-gauge or local-potential version of such treatments.

Applying the Dirac operator to the left coordinate $x_1$ and then
Wigner transforming gives a left star equation
$
H\star W=0,
$
where
$
H(\xi,p)
=
    \gamma^\mu(c p_\mu - qA_\mu(\xi))-mc^2 \, \Id.
$
Similarly, applying the adjoint Dirac equation to the right coordinate
$x_2$ gives
\mbox{$
W\star H=0
$}.
Thus, we find that the Dirac equation for the spinor amplitude becomes a pair of left and right phase-space equations for the spinor-matrix Wigner
function:
$$
H\star W=0,
\qquad
W\star H=0.
$$
These equations are consistent with the left and right zero-eigenvalue equations in Eq. \eqref{eq:two-sided-zero-constraint}, with $\epsilon =0$ on the mass shell.
Thus the matrix theory appears to be in agreement with the phase-space Dirac equation. 

To complete the analogy to the previous section, taking the difference of the two Dirac-Wigner equations gives the matrix Moyal-bracket transport equation
$$
H\star W-W\star H=0,
$$
or $\MB{H}{W}=0$, the usual transport equation. Their sum,
$
H\star W+W\star H=0,
$ supplies the companion constraint equation.


\section{The minimal argument}
\label{sec:minimal-argument}
The argument presented so far has been arranged to explain the motivation for each step in the construction. 
This order, however, makes the assumptions needed to reach the final result less immediately apparent as some conditions introduced as ansatz are subsequently recovered as consequences of later requirements. 
We therefore recast the argument in a more direct order, distinguishing the required assumptions from the conditions they imply and omitting those that need not be imposed separately. 
This gives a minimal formulation of the paper’s core thesis that quantum phase-space mechanics can emerge from a classical relativistic statistical starting point that retains both sheets of the massive mass shell, under the below stated factorisation, deformation and spectral requirements.

Begin on flat canonical phase spacetime (Ansatz~\ref{ans:flat-phase-spacetime}) with a massive particle, retaining both sheets $p^\mu_\sigma p_{\sigma\mu}=m^2c^2$, $\sigma=\pm1$. 
As in Ansatz~\ref{ans:complete-mass-shell-factorisation}, require a single finite-dimensional factor $K(p)=A^\mu p_\mu$, linear in all four momentum components, with $K(p)^2=p^\mu p_\mu\Id$ for arbitrary $p$. 
Taking the minimal complex representation of Ansatz~\ref{ans:minimal-clifford-representation} gives
\begin{equation}
 \COM{A^\mu}{A^\nu}_+=2\eta^{\mu\nu}\Id,\qquad
 N=4,\qquad
 \operatorname{rank}\Pi_+(p_\sigma)=2.
 \label{eq:minimal-clifford}
\end{equation}
Statistical completeness within the selected kernel (Ansatz~\ref{ans:statistical-completeness}) requires
 $\uW_0=\Pi_+\uW_0\Pi_+$, with an arbitrary $\uF_\sigma\in M_2(\mathbb C)$ acting on that kernel. 
The spinor structure is therefore present before a quantum scale has been introduced.

To expose the higher members of the preservation hierarchy, take the smooth commuting covector $\kappa_\mu(x,p)$ of Ansatz~\ref{ans:phase-space-generator}, with $\uH_\kappa=\gamma^\mu\kappa_\mu-mc^2\Id$. Work near a regular mass shell in the timelike sector (Ansatz~\ref{ans:timelike-sector}), so that the selected projector is well defined. 
The on-shell density is represented on the full phase space by
\begin{equation}
 \uW_0
 =\delta\!\left(\kappa^\mu\kappa_\mu-m^2c^4\right)
   \Pi_+\widetilde{\uF}\Pi_+,\qquad
 \uH_\kappa\uW_0=\Zd=\uW_0\uH_\kappa.
 \label{eq:minimal-ambient}
\end{equation}
Here $\widetilde{\uF}$ is a smooth extension of the on-shell density so products of smooth symbols with $\uW_0$ are understood distributionally, order by order.

Now take a universal local bidifferential product subject to the structural requirements of Ansatz~\ref{ans:universality}--\ref{az:bg}.
In particular, constant matrices multiply in the ordinary way, the product is invariant under translations and linear canonical transformations, and it is associative (Ansatz~\ref{ans:associativity}). 
It contains no additional phase-space tensors or independent action scales (Ansatz~\ref{az:bg}). 
We require the first-order commutator of the universal product to reproduce the classical transport and constraint-reaction condition in Eq. (40), and hence the projected Liouville equation in Eq. (39). Its Poisson-bracket limit is therefore a matching condition on the continuation of the classical dynamics.
These restrictions concern the universal product, not the generator $\uH_\kappa$. Since $\mathcal P$ has units of inverse action, $\lambda$ must carry units of action and dimensional consistency gives the order-by-order homogeneity specified in Ansatz~\ref{ans:order-homogeneity}. 
These conditions determine
\begin{equation}
 \begin{aligned}
  C_n(\uF,\uG)&=a_n\uF\mathcal P^n\uG,
      &a_n&=\frac{a_1^n}{n!},\\
  \uF\star\uG&=\uF\exp(a_1\lambda\,\mathcal P)\uG,
      &a_1&\ne0.
 \end{aligned}
 \label{eq:minimal-star}
\end{equation}
Thus the first derivative term has the Weyl-symmetrised form without requiring that ordering as a separate premise. The conditions on the product do not fix the physical value of $\lambda$ or the phase of $a_1$.

Require the deformed distribution to preserve the factor constraint at every order (Ansatz~\ref{ans:all-order-preservation}), $\uH_\kappa\star\uW_\lambda -\uW_\lambda\star\uH_\kappa=\Zd$, where $\uW_\lambda=\uW_0+\lambda\uW_1+\cdots$. 
Its first two orders give
\begin{equation}
 [\uH_\kappa,\uW_0]=\Zd,\qquad
 [\uH_\kappa,\uW_1]
 +2a_1\mathcal L^{\mathcal W}_{\uH_\kappa}(\uW_0)=\Zd.
 \label{eq:minimal-first-order}
\end{equation}
Since the diagonal projection annihilates a commutator with $\uH_\kappa$, the second equation implies $\mathsf D_\kappa[
 \mathcal L^{\mathcal W}_{\uH_\kappa}(\uW_0)]=\Zd$.
For the free generator this becomes $p^\mu_\sigma\partial_\mu\uF_\sigma/m=0_2$.
The projected free transport of Ansatz~\ref{ans:projected-classical-transport} and the classical preservation condition of Ansatz~\ref{az:classical-preservation} are thus recovered from the same all-order requirement, together
with the first-order inter-sector reaction.

Finally, retaining the selected \emph{star}-spectral sector
(Ansatz~\ref{ans:star-spectral-sector}) is an additional requirement.
For its zero eigenvalue, one left star constraint, together with commutator preservation, gives
\begin{equation}
 \uH_\kappa\star\uW_\lambda=\Zd
 =\uW_\lambda\star\uH_\kappa.
 \label{eq:minimal-two-sided}
\end{equation}
Comparison with the Dirac--Wigner equations selects the physical normalisation $a_1\lambda=\ui\hbar/2$, conventionally written $a_1=\ui/2$ and $\lambda=\hbar$ as in
Ansatz~\ref{ans:planck-constant}.
 
\begin{table*}[!th]
\centering
\caption{Role of each ansatz in the reordered minimal argument.}
\label{tab:minimal-ansatz}
\small
\setlength{\tabcolsep}{4pt}
\renewcommand{\arraystretch}{1.12}
\begin{tabular}{p{0.29\textwidth}p{0.14\textwidth}p{0.51\textwidth}}
\hline
Ansatz & Status & Place in the minimal argument \\
\hline
1. Flat phase spacetime
& Setting
& Supplies flat canonical phase space and its constant Poisson tensor. \\

2. Complete first-order mass-shell factorisation
& Required
& Retains both massive sheets and requires a finite-dimensional factor linear in all four momentum components and determines the Clifford relations. \\

3. Minimal Clifford representation
& Required
& Selects the four-dimensional complex representation and hence the rank-two on-shell kernel. \\

4. Statistical completeness within the selected sector
& Required
& Retains arbitrary internal populations and coherences within that kernel. \\

5. Projected Weyl-ordered transport
& Consequence
& Recovered for the free generator from first-order star-commutator preservation, once the first-order deformation is nonzero. \\

6. Phase space generator
& Required
& Extends the free construction to smooth generators $\kappa_\mu(x,p)$, allowing the higher preservation conditions to be examined. \\

7. Timelike sector
& Domain
& Holds on the massive shell and in a sufficiently small neighbourhood and ensures that the projectors remain well defined. \\

8. Massive particles
& Already included
& The condition $m>0$ is already part of Ansatz~2 so no separate assumption is needed. \\

9. Classical preservation condition
& Consequence
& Follows by projecting the first-order equation of Ansatz~20, with the nonzero Poisson-form first-order product. \\

10. Self-preservation of a generator
& Automatic
& Follows from using a star commutator: $\COM{F}{F}_\star=0$ (where, for us $\COM{\cdot}{\cdot}_\star= \MB{\cdot}{\cdot}$). This ansatz motivates that choice without adding a further restriction to it. \\

11. Universality
& Combined
& Retained as a requirement of the product in Ansatz~12 so need not be stated separately. \\

12. Star product
& Required
& Introduces a universal, local, unital, formal bidifferential deformation of ordinary matrix-valued multiplication. \\

13. Symplectic form
& Required
& Requires full linear symplectic covariance of the product and combines with Ansatz~14 as \emph{affine symplectic covariance}. \\

14. Translation invariance of $\star$
& Required
& Gives constant bidifferential coefficients and retained within the combined \emph{affine covariance} requirement. \\

15. Constant matrix neutrality
& Required
& Leaves multiplication by constant matrices undeformed which, with associativity, reduces the coefficients to scalar bidifferential operators acting entry by entry. \\

16. Associativity of $\star$
& Required
& Fixes the exponential coefficient relations and gives the star commutator its Jacobi and derivation properties. \\

17. No additional background structure
& Required
& Restricts the product to canonical Poisson contractions and dimensionless coefficients, without further phase-space tensors or action scales. \\

18. Order-by-order homogeneity
& Conditional consequence
& Follows dimensionally if $\lambda$ is action-valued and Ansatz~17 is retained, otherwise it remains an independent selection requirement. \\

19. Planck's constant
& Physical identification
& Fixes the physical normalisation $a_1\lambda=i\hbar/2$ by comparison with quantum theory or experiment, rather than by the structural argument. \\

20. All-order constraint preservation
& Required
& Requires the entire star-commutator hierarchy to vanish to yield classical preservation and the successive inter-sector reactions. \\

21. Preservation of the selected spectral sector
& Required
& Adds a left star-spectral constraint which, together with Ansatz~20, supplies the matching right constraint and the two-sided Dirac--Wigner form. \\
\hline
\end{tabular}
\end{table*}
\begin{table*}[!th]
\centering
\caption{Relation of each ansatz to established classical and quantum theory. Note that a familiar property or analogue need not imply the corresponding requirement on the present matrix-valued deformation.}
\label{tab:ansatz-classical-quantum}
\footnotesize
\setlength{\tabcolsep}{4pt}
\renewcommand{\arraystretch}{1.12}
\begin{tabular}{p{0.24\textwidth}p{0.35\textwidth}p{0.35\textwidth}}
\hline
Ansatz & Classical physics & Quantum theory \\
\hline
1. Flat phase spacetime
& Standard setting for special-relativistic Hamiltonian mechanics where flatness is a choice of scope.
& Standard setting for flat-space relativistic quantum theory but not a universal requirement. \\

2. Complete first-order mass-shell factorisation
& The mass shell is standard. Retaining both sheets and requiring finite-dimensional matrix factorisation add restrictions.
& Linear factorisation and both energy branches are familiar from Dirac theory. \\

3. Minimal Clifford representation
& Not required by ordinary scalar particle mechanics.
& The four-dimensional complex representation is standard for a massive Dirac particle in $3+1$ dimensions but argued from the Schr\"odinger equation. \\

4. Statistical completeness within the selected sector
& Retaining general statistics is familiar however internal matrix coherences add structure beyond scalar Liouville mechanics.
& Arbitrary populations and coherences within a specified internal sector are standard density-matrix practice. \\

5. Projected Weyl-ordered transport
& Scalar Liouville transport is standard however matrix ordering and spectral projection add structure.
& Symmetrised matrix Poisson terms occur in quantum--classical Liouville theory but the present projection is a specific construction. \\

6. Phase space generator
& Smooth generators are standard but the Clifford-valued family built from arbitrary $\kappa_\mu(x,p)$ is an extension.
& Matrix-valued generators are familiar however this particular family is not a general quantum postulate. \\

7. Timelike sector
& Standard domain for massive relativistic motion.
& Equally familiar for massive relativistic particles. \\

8. Massive particles
& Conventional choice of physical system and massless systems not yet considered.
& The same choice is conventional and massless systems not yet considered. \\

9. Classical preservation condition
& Constraint preservation is standard and the spectral-diagonal matrix condition is our particular implementation.
& Compatible with constrained quantum dynamics, but it is not a universal quantum postulate. \\

10. Self-preservation of a generator
& Standard bracket property: $\PB{H}{H}=0$.
& Standard commutator property: $\COM{H}{H}=0$. \\

11. Universality
& Ordinary multiplication and the canonical Poisson bracket do not depend on the Hamiltonian.
& Operator multiplication is likewise Hamiltonian-independent and so universal symbol products are familiar. \\

12. Star product
& Introducing a deformation of ordinary multiplication adds a composition rule beyond classical mechanics.
& Standard in deformation quantisation and phase-space quantum mechanics (recovered here) . \\

13. Linear symplectic covariance
& Standard property of the canonical Poisson bracket.
& Standard for the Weyl--Moyal product and requiring it restricts the symbol calculus. \\

14. Translation invariance of the product
& Standard for ordinary multiplication and the constant canonical Poisson bracket, even when the dynamics breaks this symmetry.
& Standard for the flat Weyl--Moyal product, even when the dynamics breaks this symmetry. \\

15. Constant matrix neutrality
& Constants have a scalar analogue for us the internal matrix requirement adds structure beyond scalar mechanics.
& Standard for orbital symbols combined with an undeformed internal matrix algebra but is not compulsory in every quantisation scheme (alternatives a subject for further work). \\

16. Associativity
& Ordinary multiplication is associative and the Poisson bracket separately satisfies Jacobi identity.
& Operator multiplication and star products are associative. \\

17. No additional background structure
& The canonical Poisson bracket uses no further phase-space tensors but extending this restriction to all deformation orders adds a requirement.
& Satisfied by the standard flat Moyal construction, but not universally required of quantum theory. \\

18. Order-by-order homogeneity
& No counterpart in ordinary Liouville mechanics, which has no deformation series.
& Holds for the standard Moyal expansion but is not required of every equivalent star-product representation (alternatives a subject for further work). \\

19. Planck's constant
& Not supplied by classical mechanics (outside of our construction which follows classical arguments).
& Standard empirical normalisation of quantum theory. \\

20. All-order constraint preservation
& Successive consistency conditions are familiar from classical constraint algorithms here the present deformation hierarchy adds structure.
& Commutator-based preservation is familiar and vanishing of this particular hierarchy constrains the admissible distributions (alternatives a subject for further work). \\

21. Preservation of the selected spectral sector
& Restriction to a constraint surface provides an analogue, without an ordinary scalar star-spectral formulation.
& Two-sided star-eigenvalue equations are standard within a specified eigenvalue sector for us continuation of the selected sector remains an additional requirement (alternatives a subject for further work). \\
\hline
\end{tabular}
\end{table*}
 
This ordering of the argument clarifies which requirements are independent and which are consequences of the construction. For clarity a summary of the status of each ansatz in this minimal argument if given in Table~\ref{tab:minimal-ansatz}.
Also in Table~\ref{tab:ansatz-classical-quantum} to make clear what new assumptions and their combinations are made in this work we describe the relation of each ansatz to established classical and quantum theory.
For a nonzero first-order deformation, all-order preservation recovers both the projected classical transport and the first-order compatibility condition, together with the complementary reaction required to preserve the constraint. 
These therefore need not be imposed separately in the minimal argument. 
The restrictions on the universal product and the continuation of the selected star-spectral sector remain explicit assumptions. 
Within this class, the construction selects the Weyl–Moyal composition law and the two-sided Dirac–Wigner equations. 
Complete mass-shell factorisation supplies the spinor structure before an action scale is introduced and finally comparison with quantum theory or experiment fixes the physical normalisation.


\section{Concluding remarks}
This work separates two ingredients that are normally introduced together in the Dirac theory: the spinor algebra and its quantum scale. 
A finite-dimensional factorisation of the complete massive mass-shell constraint, linear in all four components of four-momentum, necessarily generates a Clifford algebra. 
In $3+1$ dimensions its minimal complex representation is four-dimensional, whereas the on-shell kernel of either factor has rank two. 
Statistical completeness within that kernel is assumed and is represented by an arbitrary $2\times2$ matrix distribution.
This structure arises before any operator substitution or introduction of a constant with dimensions of action.
Projecting the Weyl-ordered matrix Liouvillian onto the selected rank-two sector yields
$$
    \PD{\uF_\sigma}{t}
    +
    \sigma\frac{c^2p^i}{E_p}\PD{\uF_\sigma}{x^i}
    =0_2,
$$
and therefore transports arbitrary internal populations and coherences with the appropriate relativistic branch velocity. 
The complementary projector blocks should not be imposed as further classical transport equations. 
Instead, they represent the reaction required to preserve the spectral constraint and determine the inter-sector correction at the next order of the deformation.

For a universal, local and unital deformation on flat canonical phase space, affine symplectic covariance, constant-matrix neutrality, absence of additional background structure, order-by-order homogeneity and associativity select the Weyl–Moyal exponential up to the normalisation of the deformation parameter. Its first-order star commutator reproduces the symmetrised matrix Liouvillian used provisionally in Part I. For a nonzero first-order deformation, all-order constraint preservation then recovers the projected classical transport and the first-order compatibility condition, showing that these earlier requirements are consequences within the completed construction.

Dimensional consistency requires the deformation parameter $\lambda$ to have units of action. 
Its numerical identification with $\hbar$ requires comparison with established quantum mechanics or experiment as this cannot follow from the classical assumptions alone. 
This identification supplies the angular-momentum scale that converts the dimensionless two-state rotation algebra into physical spin-$\tfrac12$.

Requiring the complete constraint-preservation hierarchy is equivalent to the vanishing matrix star commutator. With the additional requirement that the deformed distribution remain in the selected star-spectral sector, the mass-shell constraint becomes
$$
\uH_\kappa\star \uW=\Zd=\uW\star \uH_\kappa,
$$
which has the same form as the left and right Dirac–Wigner equations. 
The central conclusion is therefore that relativistic mass-shell factorisation supplies spinor structure before quantisation, while completion of the constraint-preservation hierarchy within the stated deformation class leads to the Weyl–Moyal composition law and, with the additional spectral condition, the Dirac–Wigner phase-space equations. 
The classical starting point therefore includes the finite-dimensional factorisation, linear in all four momentum components and squaring to the mass-shell quadratic form,
specified in Ansatz~\ref{ans:complete-mass-shell-factorisation}.
This factorisation requirement, motivated by retaining both sheets, implies the Clifford algebra. 
The Weyl–Moyal and Dirac–Wigner conclusions additionally depend on the stated deformation and spectral assumptions.
The physical action scale enters via $\lambda$ which is identified with $\hbar$ by comparison with quantum theory or experiment.

Finally we note that the Clifford-algebra result is local and does not itself require globally flat spacetime. We speculate that the exact global Moyal form does: on curved phase space one should expect a symplectic connection and curvature-dependent corrections. Extensions to massless sectors, more general Clifford-valued generators and fully gauge-covariant Wigner functions provide natural tests of the scope of the construction beyond those explored here.

The central result is that a classically formulated requirement on relativistic statistical evolution leads, through its consistent completion, to quantum dynamics. 
Motivated by retaining both sheets of the mass shell, the construction begins with transport within the sector selected by mass-shell factorisation and the reaction required to preserve that constraint. 
Requiring a universal associative completion that retains this classical transport, together with the stated symmetry and spectral conditions, selects the Weyl–Moyal composition law and the Dirac–Wigner equations. 
Through the Weyl–Wigner correspondence, the result also admits its equivalent operator formulation. The construction separates the origin of spinor structure, the selection of quantum composition and the empirical determination of the physical action scale. 
Quantum structure thus emerges without postulating quantum operators or a specific quantisation rule, rather it is selected by the requirements for completing the classical statistical dynamics with this framework.

\section*{Acknowledgements}
For detailed comments and pertinent questions on this version I am most thankful to Sasha Balanov and Bill Munro.
I also am grateful to those who commented on earlier versions of this work. 
In particular, I am indebted to Tim Spiller and Raymond Bishop for detailed feedback and incisive questions, especially to Tim for his challenge on how to interpret the classical limit of spin in the absence of Planck's constant and Ray for his reviews and suggestions on later drafts of the manuscript. 
I thank John Samson for repeatedly identifying mistakes in early drafts.
I also thank Todd Tilma, Jason Ralph, Russell Rundle, and Kieran Bjergstrom for constructive discussions that helped test the novelty of the ideas and clarify their presentation, and Alexander Balanov and Alexandre Zagoskin for additional stimulating conversations. 
Ben Davies kindly suggested improvements that enhanced the readability of the manuscript.

\section*{CRediT authorship contribution statement}
\textbf{Mark Everitt:} Conceptualisation, Methodology, Formal analysis,
Investigation, Writing -- original draft, Writing -- review \& editing.

\section*{Funding}
This research did not receive any specific grant from funding agencies in the public, commercial, or not-for-profit sectors.

\section*{Declaration of competing interest}
The author declares that he has no known competing financial interests or personal relationships that could have appeared to influence the work reported in this paper.

\section*{Data availability}
No data were generated or analysed for this theoretical study.

\section*{Declaration of generative AI use}
The author acknowledges the use of successive versions of OpenAI's ChatGPT during the development of this work for checking algebraic consistency, exploring alternative formulations, identifying possible gaps in reasoning, and improving the clarity of the presentation. 
The author reviewed and edited the resulting material and takes full responsibility for the content of the article.

\bibliographystyle{elsarticle-num}
\bibliography{ref}

@book{GoodmanWallach2009,
  author    = {Roe Goodman and Nolan R. Wallach},
  title     = {Symmetry, Representations, and Invariants},
  series    = {Graduate Texts in Mathematics},
  volume    = {255},
  publisher = {Springer},
  address   = {New York},
  year      = {2009},
  doi       = {10.1007/978-0-387-79852-3}
}

@book{EtingofEtAl2011,
  author    = {Etingof, Pavel and Golberg, Oleg and Hensel, Sebastian
               and Liu, Tiankai and Schwendner, Alex and Vaintrob, Dmitry
               and Yudovina, Elena},
  title     = {Introduction to Representation Theory},
  series    = {Student Mathematical Library},
  volume    = {59},
  publisher = {American Mathematical Society},
  address   = {Providence, RI},
  year      = {2011},
  doi       = {10.1090/stml/059}
}

@article{BRAVYI20112793,
title = {Schrieffer–Wolff transformation for quantum many-body systems},
journal = {Annals of Physics},
volume = {326},
number = {10},
pages = {2793-2826},
year = {2011},
issn = {0003-4916},
doi = {https://doi.org/10.1016/j.aop.2011.06.004},
url = {https://www.sciencedirect.com/science/article/pii/S0003491611001059},
author = {Sergey Bravyi and David P. DiVincenzo and Daniel Loss}
}

@article{BAYEN197861,
title = {Deformation theory and quantization. I. Deformations of symplectic structures},
journal = {Annals of Physics},
volume = {111},
number = {1},
pages = {61-110},
year = {1978},
issn = {0003-4916},
doi = {https://doi.org/10.1016/0003-4916(78)90224-5},
url = {https://www.sciencedirect.com/science/article/pii/0003491678902245},
author = {F Bayen and M Flato and C Fronsdal and A Lichnerowicz and D Sternheimer}
}

@article{gutt_natural_2003,
	title = {Natural {Star} {Products} on {Symplectic} {Manifolds} and {Quantum} {Moment} {Maps}},
	volume = {66},
	issn = {1573-0530},
	url = {https://doi.org/10.1023/B:MATH.0000017717.51035.f1},
	doi = {10.1023/B:MATH.0000017717.51035.f1},
	language = {en},
	number = {1},
	urldate = {2026-09-18},
	journal = {Letters in Mathematical Physics},
	author = {Gutt, Simone and Rawnsley, John},
	month = oct,
	year = {2003},
	pages = {123--139},
}

@article{Howe1989,
 ISSN = {00029947},
 URL = {http://www.jstor.org/stable/2001418},
 author = {Roger Howe},
 journal = {Transactions of the American Mathematical Society},
 number = {2},
 pages = {539--570},
 publisher = {American Mathematical Society},
 title = {Remarks on Classical Invariant Theory},
 urldate = {2026-09-18},
 volume = {313},
 year = {1989}
}

@misc{everitt2026massshellfactorisationspinattempt,
      title={From Mass-Shell Factorisation to Spin: An Attempt at a Matrix-Valued Liouville Framework for Relativistic Classical and Quantum Phase-Spacetime}, 
      author={Mark J. Everitt},
      year={2026},
      eprint={2505.03551},
      archivePrefix={arXiv},
      primaryClass={quant-ph},
      url={https://arxiv.org/abs/2505.03551}, 
}

@article{Dirac28,
    author = {Dirac, Paul Adrien Maurice},
    title = {The quantum theory of the electron},
    journal = {Proceedings of the Royal Society of London. Series A, Containing Papers of a Mathematical and Physical Character},
    volume = {117},
    number = {778},
    pages = {610-624},
    year = {1928},
    month = {02},
    issn = {0950-1207},
    doi = {10.1098/rspa.1928.0023},
    url = {https://doi.org/10.1098/rspa.1928.0023},
    eprint = {https://royalsocietypublishing.org/rspa/article-pdf/117/778/610/25048/rspa.1928.0023.pdf},
}

@Article{sym18061018,
AUTHOR = {Ji, Chueng-Ryong and Piasecki, Daniel W.},
TITLE = {Origin of the Covariant Wigner Operator as a Quantum Amplitude in QCD},
JOURNAL = {Symmetry},
VOLUME = {18},
YEAR = {2026},
NUMBER = {6},
ARTICLE-NUMBER = {1018},
URL = {https://www.mdpi.com/2073-8994/18/6/1018},
ISSN = {2073-8994},
DOI = {10.3390/sym18061018}
}

@article{Moyal_1949, 
title={Quantum mechanics as a statistical theory}, 
volume={45}, 
DOI={10.1017/S0305004100000487}, 
number={1}, 
journal={Mathematical Proceedings of the Cambridge Philosophical Society}, 
author={Moyal, J. E.}, year={1949}, pages={99–124}}

@article{GROENEWOLD1946405,
title = {On the principles of elementary quantum mechanics},
journal = {Physica},
volume = {12},
number = {7},
pages = {405-460},
year = {1946},
issn = {0031-8914},
doi = {https://doi.org/10.1016/S0031-8914(46)80059-4},
url = {https://www.sciencedirect.com/science/article/pii/S0031891446800594},
author = {H.J. Groenewold}
}

@article{wilsonLine,
author = {Gao, Jian-Hua and Liang, Zuo-Tang and Wang, Qun},
title = {Quantum kinetic theory for spin-1/2 fermions in Wigner function formalism},
journal = {International Journal of Modern Physics A},
volume = {36},
number = {01},
pages = {2130001},
year = {2021},
doi = {10.1142/S0217751X21300015},
URL = {https://doi.org/10.1142/S0217751X21300015},
eprint = {https://doi.org/10.1142/S0217751X21300015}
}

@book{thaller1992dirac,
  title     = {The Dirac Equation},
  author    = {Thaller, Bernd},
  publisher = {Springer},
  year      = {1992}
}

@book{BjorkenDrell1964RQM,
  author    = {Bjorken, James D. and Drell, Sidney D.},
  title     = {Relativistic Quantum Mechanics},
  publisher = {McGraw-Hill},
  year      = {1964},
  note      = {See Section 3.2, ``Projection Operators for Energy and Spin'', Eq. (3.18).}
}

@article{D1,
  title = {Kinetic theory for massive $\mathrm{spin}\text{\ensuremath{-}}1/2$ particles from the Wigner-function formalism},
  author = {Weickgenannt, Nora and Sheng, Xin-li and Speranza, Enrico and Wang, Qun and Rischke, Dirk H.},
  journal = {Phys. Rev. D},
  volume = {100},
  issue = {5},
  pages = {056018},
  numpages = {18},
  year = {2019},
  month = {Sep},
  publisher = {American Physical Society},
  doi = {10.1103/PhysRevD.100.056018},
  url = {https://link.aps.org/doi/10.1103/PhysRevD.100.056018}
}

@article{D2,
title = {Transport equations for the QCD quark Wigner operator},
journal = {Nuclear Physics B},
volume = {276},
number = {3},
pages = {706-728},
year = {1986},
issn = {0550-3213},
doi = {https://doi.org/10.1016/0550-3213(86)90072-6},
url = {https://www.sciencedirect.com/science/article/pii/0550321386900726},
author = {H.-Th. Elze and M. Gyulassy and D. Vasak}
}

@article{D3,
title = {Relativistic Quantum Transport Theory for Electrodynamics},
journal = {Annals of Physics},
volume = {245},
number = {2},
pages = {311-338},
year = {1996},
issn = {0003-4916},
doi = {https://doi.org/10.1006/aphy.1996.0011},
url = {https://www.sciencedirect.com/science/article/pii/S0003491696900111},
author = {P. Zhuang and U. Heinz}
}

@article{D4,
title = {Quantum transport theory for abelian plasmas},
journal = {Annals of Physics},
volume = {173},
number = {2},
pages = {462-492},
year = {1987},
issn = {0003-4916},
doi = {https://doi.org/10.1016/0003-4916(87)90169-2},
url = {https://www.sciencedirect.com/science/article/pii/0003491687901692},
author = {David Vasak and Miklos Gyulassy and Hans-Thomas Elze}
}

@article{MARSDEN198629,
title = {Covariant poisson brackets for classical fields},
journal = {Annals of Physics},
volume = {169},
number = {1},
pages = {29-47},
year = {1986},
issn = {0003-4916},
doi = {https://doi.org/10.1016/0003-4916(86)90157-0},
author = {J.E Marsden and R Montgomery and P.J Morrison and W.B Thompson}
}

@book{hakim2011introduction,
  title={Introduction to Relativistic Statistical Mechanics: Classical and Quantum},
  author={Hakim, R.},
  isbn={9789814322430},
  lccn={2010054042},
  series={G - Reference,Information and Interdisciplinary Subjects Series},
  year={2011},
  publisher={World Scientific}
}

@article{RevModPhys.35.350,
  title = {Relativistic Invariance and Hamiltonian Theories of Interacting Particles},
  author = {Currie, D. G. and Jordan, T. F. and Sudarshan, E. C. G.},
  journal = {Rev. Mod. Phys.},
  volume = {35},
  issue = {2},
  pages = {350--375},
  numpages = {0},
  year = {1963},
  month = {Apr},
  publisher = {American Physical Society},
  doi = {10.1103/RevModPhys.35.350},
  url = {https://link.aps.org/doi/10.1103/RevModPhys.35.350}
}

@article{PhysRevA.99.012115,
  title = {General approach to quantum mechanics as a statistical theory},
  author = {Rundle, R. P. and Tilma, Todd and Samson, J. H. and Dwyer, V. M. and Bishop, R. F. and Everitt, M. J.},
  journal = {Phys. Rev. A},
  volume = {99},
  issue = {1},
  pages = {012115},
  numpages = {13},
  year = {2019},
  month = {Jan},
  publisher = {American Physical Society},
  doi = {10.1103/PhysRevA.99.012115},
  url = {https://link.aps.org/doi/10.1103/PhysRevA.99.012115}
}
\end{document}